\documentclass[a4paper,aps,prd,twocolumn,eqsecnum,amsmath,amssymb,nofootinbib,superscriptaddress]{revtex4-2}
\usepackage{dcolumn}
\usepackage{bm}
\usepackage{graphics}
\usepackage{graphicx}
\usepackage{epsfig}
\usepackage{booktabs,multirow}
\usepackage{hhline}
\usepackage{slashed}
\usepackage{rccol}
\usepackage{mathrsfs}
\usepackage[dvipsnames]{xcolor,colortbl}

\usepackage{xcolor}\colorlet{linkequation}{blue}
\usepackage[colorlinks=true, 
            linkcolor=red,   
            filecolor=red,   
            citecolor=green, 
            urlcolor=blue,
            hyperfootnotes=true]{hyperref}

\newcommand*{\SavedEqref}{}
\let\SavedEqref\eqref
\renewcommand*{\eqref}[1]{%
  \begingroup
    \hypersetup{
     linkcolor=linkequation,
      linkbordercolor=linkequation,
    }%
    \SavedEqref{#1}%
  \endgroup
}

\begin{document}

\def\beqa{\begin{eqnarray}}
\def\eeqa{\end{eqnarray}}
\newcommand{\be}{\ensuremath{\beta}}
\newcommand{\al}{\ensuremath{\alpha}}
\newcommand{\sa}{\ensuremath{\sin\alpha}}
\newcommand{\ca}{\ensuremath{\cos\alpha}}
\newcommand{\ta}{\ensuremath{\tan\alpha}}
\newcommand{\sbt}{\ensuremath{\sin\beta}}
\newcommand{\cbt}{\ensuremath{\cos\beta}}
\newcommand{\cba}{\ensuremath{c_{\beta-\alpha}}}
\newcommand{\ma}{\ensuremath{m_{A}}}
\newcommand{\mh}{\ensuremath{m_{h^0}}}
\newcommand{\mH}{\ensuremath{m_{H^0}}}
\newcommand{\mev}{\mbox{~MeV}}
\newcommand{\gev}{\mbox{~GeV}}
\newcommand{\tev}{\mbox{~TeV}}

\newcommand{\ben}{\begin{enumerate}}
\newcommand{\een}{\end{enumerate}}
\newcommand{\bc}{\begin{center}}
\newcommand{\ec}{\end{center}}
\newcommand{\mb}{\mbox{\ }}
\newcommand{\vs}{\vspace}
\newcommand{\ra}{\rightarrow}
\newcommand{\la}{\leftarrow}
\newcommand{\ul}{\underline}
\newcommand{\ds}{\displaystyle}
\definecolor{LightCyan}{rgb}{0.0, 1, 0.94}

\title{Beam polarization effects on Z-boson pair production at electron-positron colliders: a full one-loop analysis}
\author{Mehmet~Demirci}
\email{mehmetdemirci@ktu.edu.tr}
\affiliation{Department of Physics, University of Wisconsin–Madison, Madison, Wisconsin 53706, USA}
\affiliation{Department of Physics, Karadeniz Technical University, Trabzon 61080, Turkey}%
\author{A.~Baha~Balantekin}
\email{baha@physics.wisc.edu}
\affiliation{Department of Physics, University of Wisconsin–Madison, Madison, Wisconsin 53706, USA}

\date{\today}
\begin{abstract} We present high-precision predictions for Z boson-pair production via electron-positron collisions by taking into account a full set of one-loop order scattering amplitudes, i.e., electroweak (EW) corrections together with soft and hard QED radiation.
We provide a detailed numerical evaluation, from full EW corrections to pure QED corrections, specifically focusing on the effect of initial beam polarization on production rate. The left-right asymmetry and angular distributions are also presented. The radiative corrections are largely affected by initial beam polarizations. We get an improvement as around three times with completely polarized beams of $e^-_L$ and $e^+_R$. We find that the radiative corrections can sizably modify the production rate, typically yielding a total relative correction up to a few tens of a percent. This imply that the full EW corrections are required for $e^- e^+ \to ZZ$ to match with percent level accuracy.
\keywords{Electroweak radiative corrections, Z boson, Beam polarization, Standard Model, electron-positron collider}
\end{abstract}

\maketitle

\section{Introduction}
The Standard Model (SM)~\cite{Glashow61,Weinberg67,Salam68} has been well established as a self-consistent gauge theory. Starting with the discovery of the Higgs boson~\cite{ATLAS,CMS} a decade ago, we have an increasing confidence on the consistency of the Standard Model (SM) from the LHC experiments ~\cite{ATLASCMS}. Despite its many successes, there are remaining questions (such as the hierarchy problem, neutrino masses, dark matter, the strong CP problem, and the generation of a baryon asymmetry) which require beyond the SM (BSM) physics. Current and future experiments will continue not only make sensitivity tests of the SM, but also will search for BSM physics to meet the challenges listed above.

Proton-proton colliders have significant backgrounds which makes it harder to explore rare processes. On the other hand, electron-positron colliders have cleaner backgrounds and offer important opportunities to precisely observe the interesting phenomena. Various proposals for such colliders exist: The International Linear Collider (ILC)~\cite{ILC1,ILC2,ILC3} is one of the proposed linear colliders, which would be operated in the c.m. energies of $\sqrt{s}=250-500\gev$ (extendable up to a 1 TeV) with collision modes of $e^-e^+$, $e^-e^-$, $e^-\gamma$, and $\gamma\gamma$. At the ILC, it is also expected to use the electron and positron beams with $80\%$ and $30\%$ polarizations, respectively. Another linear collider is the Compact Linear Collider (CLIC)~\cite{CLIC1}, which is foreseen to be operated a TeV scale high-luminosity capacity, in stages at energies of $\sqrt{s}=380\gev$, $1.5\tev$ and $3\tev$. Furthermore, there are circular collider projects such as Circular Electron-Positron Collider (CEPC)~\cite{CEPC}, and Future Circular Collider
(FCC)~\cite{FCC19}. The CEPC is projected with a circumference of 100 km and to have a maximum of $\sqrt{s}=240\gev$. The FCC project~\cite{FCC19}, hosted in a 100 km tunnel at CERN, is considering three collision types: FCC-ee, for electron-positron collisions, FCC-hh, for hadron-hadron collisions, and FCC-eh, for electron-hadron collisions. The FCC-ee~\cite{FCCee}, running at different energies to research the production of Z, W, H and $t\bar{t}$, is expected to be a Higgs, electroweak and top factory with high-luminosity. The above lepton colliders are expected to offer the unique opportunities for precise measurements and signs of new physics, appearing in the form of small deviations from the SM.

In parallel, high-precision predictions from theory are required to get more precise measurements for clues of the BSM physics or more sensitivity tests of the SM.
This means that it is necessary to go beyond the lowest-order calculations for the most production channels. At least, one-loop order corrections in the production channels must be included to get adequate accuracy. However, the extent to which higher-order computations beyond one-loop order will be needed depends largely on the expected experimental accuracy. In the present study, we focus on the complete one-loop order EW corrections and beam polarization effects. The beam polarization has the ability to improve the corresponding signal-to-background ratio along with the sensitivity of the observables~\cite{Pankov06,Moortgat08,Osland10}. Furthermore, the CP-violating couplings can be separated from CP-conserving ones using beam polarization~\cite{Moortgat08}.

The main production reaction of the $Z$ boson pairs at the $e^-e^+$ colliders are $e^- e^+ \rightarrow ZZ$, which include the Born-level contributions from electron exchanges via $t$ and $u-$ channels. Since there are no Yukawa or QCD contributions, the identification of the EW corrections would be very clean. The process at Born order was studied by~\cite{Brown79,Gaemers79} a long ago. Next, one-loop EW radiative corrections were investigated by~\cite{Denner88}. However, their results directly depend on the soft cut-off parameter, since the hard photon correction is not taken into account. In addition to the above work, in Ref.~\cite{Gounaris03}, the high energy behavior of the various helicity amplitudes was investigated considering all of the MSSM contributions.

Furthermore, neutral gauge boson production via electron-positron collisions have received a significant experimental~\cite{OPAL,L3,Exp03} and theoretical~\cite{Jadach97, Gounaris00,Rahaman17,inan21,Spor22} interest, motivated by the search for anomalous neutral gauge boson self couplings. In particular, $Z$-pair events provide an opportunity for the investigation of possible triple neutral gauge boson couplings, $ZZ\gamma$ and $ZZZ$~\cite{Gounaris00_2,Alcaraz00,ATLAS13}, not allowed at tree level in the SM.

In the present work, we examine the $Z$ boson pair production via electron-positron collisions in the SM, including a full set of one-loop order EW radiative corrections, i.e., the EW corrections together with soft and hard real photon emission. We provide a detailed numerical discussion with particular emphasis on the effects of initial beam polarization. Moreover, we consider the decomposition of the EW corrections into the pure QED photonic corrections along with the corresponding counterterms and the remaining weak corrections. The left-right asymmetry and angular distributions are also presented at the Born and one-loop orders.

The rest of this paper is organized as follows. In Sec.~\ref{sec:cros}, we provide some useful analytical expressions together with the relevant Feynman diagrams and amplitudes. General shapes of the real photon emission and the virtual corrections are also discussed. In Sec.~\ref{sec:pol}, we give briefly information on initial beam polarization.  In Sec.~\ref{sec:inputs} we give a set of input parameters used in this study. In Sec.~\ref{sec:results}, we present a detailed analysis of numerical results. A comparison with the results of other approachescis also given. Finally, in Sec.~\ref{sec:conc}, we present summary and conclusions.

\section{Theoretical Setup for a cross section}\label{sec:cros}
We first present our notation regarding the Born amplitudes and one-loop EW contributions. The relevant production process is expressed as
\begin{equation} \label{eq:eeZZ}
e^{+}(p_1,\sigma_1)e^{-}(p_2,\sigma_2)\rightarrow Z(k_1,\lambda_1) Z(k_2,\lambda_2),
\end{equation}
where $\sigma_1$, $\sigma_2$, $\lambda_1$ and $\lambda_2$ denote the helicities of initial positron and electron and outgoing $Z$ bosons, respectively.
The helicities take the values $\lambda_{1,2}=0,\pm1$ and $\sigma_{1,2}=\pm1/2$. Polarization vectors of outgoing $Z$-bosons  are denoted by $\varepsilon_{\mu}(k_1,\lambda_{1})$ and $\varepsilon_{\nu}(k_2, \lambda_{2})$.
Neglecting the electron mass, the momenta in the center of mass of the initial state system are given by
\begin{equation}
\begin{split}
p_1&=\frac{\sqrt{s}}{2}(1,0,0,-1),\\
p_2&=\frac{\sqrt{s}}{2}(1,0,0,+1),\\
k_1&=\frac{\sqrt{s}}{2}(1,-\kappa \sin\theta,0,-\kappa \cos\theta),\\
k_2&=\frac{\sqrt{s}}{2}(1,+\kappa \sin\theta,0,+\kappa \cos\theta),
\end{split} \label{eq:cms_momentum}
\end{equation}
where
\begin{equation}
\kappa=\sqrt{1-\frac{4 M_Z^2}{s}},
\label{eq:kappa}
\end{equation}
and $\sqrt{s}$ and $\theta$ denotes the center of mass energy and the scattering angle, respectively. For further use, we also note the Mandelstam variables:
\begin{equation}
\begin{split}
s&=(p_1+p_2)^2=(k_1+k_2)^2,\\
t&=(k_1-p_1)^2=(k_2-p_2)^2,\\
u&=(k_2-p_1)^2=(k_1-p_2)^2.
\end{split} \label{eq:manvar}
\end{equation}
The Z-boson polarization vectors are given by
\begin{equation}
\begin{split}
\varepsilon^{\mu}_+(k,\lambda=+1)&=-\frac{1}{\sqrt{2}}(0,1,i,0),\\
\varepsilon^{\mu}_0(k,\lambda=0)&=\frac{1}{M_Z}(k_z,0,0,E),\\
\varepsilon^{\mu}_-(k,\lambda=-1)&=\frac{1}{\sqrt{2}}(0,1,-i,0),
\end{split}
\end{equation}
which obey the condition $\varepsilon^{\mu} k_\mu=0$. The sum over physical polarization states of Z-boson is given by
\begin{equation}
\sum_\lambda \varepsilon_{\mu}(k,\lambda) \varepsilon^{*}_{\nu}(k,\lambda)=-g_{\mu \nu}+\frac{k_\mu k_\nu}{M_Z^2}.
\end{equation}

The analytical and numerical evaluations are carried out by using the following tools\footnote{
Using the same tools, many calculations have been carried out with significant results (see, e.g., by one of us  ~\cite{Demirci19b,Demirci16,Demirci20,Demirci21}, and others~\cite{Juan08,Boudjema10,Heinemeyer16a,Heinemeyer18,HeYi22}).}.
The Feynman diagrams and amplitudes are created by using \textsc{FeynArts}~\cite{Feynarts1,Feynarts2}. The algebraic evaluation for amplitudes is provided technically in the same way as defined in Ref.~\cite{Demirci21}. Then, the squaring of amplitudes, the simplifying of fermion chains and the numerical computation are carried out by using \textsc{FormCalc} \cite{loop}. The scalar loop integrals are calculated via \textsc{LoopTools}~\cite{loop}. The phase-space integrations are calculated by using the Monte Carlo integration algorithm Vegas, implemented in the \textsc{CUBA} library~\cite{CUBA}. We have checked the cross sections of Born and hard photon radiation processes against the results obtained by using \textsc{Whizard}~\cite{Whizard,Omega} and \textsc{CalcHEP}~\cite{CalcHep}.

\subsection{Born amplitudes}
\begin{figure}[hbt]
    \begin{center}
\includegraphics[width=0.80\linewidth]{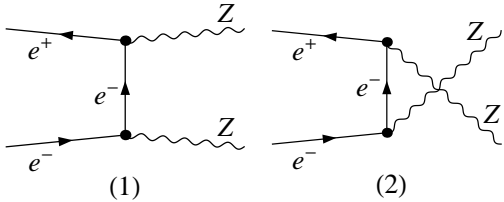}
     \end{center}
          \vspace{-4.5mm}
\caption{The lowest order Feynman diagrams contributing to $e^{-}e^{+} \rightarrow Z Z$.}\label{fig:borndiagram}
\end{figure}

At lowest order, the process $e^{+}e^{-} \rightarrow ZZ$ includes the leading contributions from $t$ and $u$-channel electron-exchange diagrams. The corresponding Feynman diagrams are given in Fig.~\ref{fig:borndiagram}, where a diagram with a Higgs field coupling to electrons is omitted. It is suppressed by a factor $m_e/\sqrt{s}$ and thus can be neglected.
The explicit expressions are given for the Born matrix elements from these diagrams as follows:
\begin{eqnarray}
\begin{split}
{\cal M}_1&=\frac{-i g^2}{4\big[t-m_{e}^2\big] c_W^2}\overline{v}(p_1,m_e) \slashed{\varepsilon}^*(k_1) \big[c_V^{e} -c_A^{e} \gamma_5 \big] \\
&\times (\slashed{p}_2-\slashed{k}_2+m_e) \slashed{\varepsilon}^*(k_2)\big[c_V^{e} -c_A^{e} \gamma_5 \big]  u(p_2,m_e),
\end{split}
\end{eqnarray}

\begin{eqnarray}
\begin{split}
{\cal M}_2&=\frac{-i g^2}{4\big[u-m_{e}^2\big] c_W^2}\overline{v}(p_1,m_e) \slashed{\varepsilon}^*(k_2) \big[c_V^{e} -c_A^{e} \gamma_5\big] \\
&\times (\slashed{p}_2-\slashed{k}_1+m_e) \slashed{\varepsilon}^*(k_1)\big[c_V^{e} -c_A^{e} \gamma_5 \big]  u(p_2,m_e),
\end{split}
\end{eqnarray}
where $g=e/s_W$, $s_W=\sin\theta_W$ and $c_W=\cos\theta_W=M_W/M_Z$. The vector and axial vector couplings are defined as $c_V^{f}=I_3^f-2Q_f s_W^2$ and $c_A^{f}=I_3^f$ for the fermion type $f$. For electron, these are $c_V^{e}=-1/2+2s_W^2$ and $c_A^{e}=-1/2$.
The Born-level total amplitude is
\begin{equation}\label{eq:totMee}
{\cal M}_{\text{Born}}=\sum_{i=1}^{2} {\cal M}_{i}.
\end{equation}
For arbitrary polarizations of the leptons and bosons, the differential cross section reads
\begin{eqnarray} \label{eq:difsigma}
\begin{split}
\biggl(\frac{d\sigma}{d\Omega }\biggr)_{\text{Born}}^{e^{+}e^{-} \rightarrow ZZ } = &\frac
{\kappa}{64 \pi^2 s} \sum_{\sigma_{1,2},\lambda_{1,2}}\frac{1}{4} (1+2 \sigma_1 P_1)(1+2 \sigma_2 P_2)\\
& \times |{\cal M}_{\text{Born}}^{\sigma_1,\sigma_2, \lambda_1,\lambda_2}|^2,
\end{split}
\end{eqnarray}
where $P_1$ and $P_2$ are the polarization degrees of the incoming electron and positron. The sum runs over all included boson polarizations. \begin{widetext}
Following the square of total amplitude and the summation over final helicities, we have
\begin{equation} \label{eq:difsigma}
\begin{split}
\biggl(\frac{d\sigma}{d t }\biggr)_{\text{Born}}^{e^{-}e^{+} \rightarrow ZZ } = &\frac{g^4}{128 \pi^2 s^2} \frac{(c_V^{e})^4+(c_A^{e})^4+6 (c_V^{e})^2 (c_A^{e})^2}{c_W^4} \biggl[ \frac{u}{t}+\frac{t}{u}+4M_Z^2 \biggl( \frac{s}{ut}\biggr) -M_Z^4\biggl(\frac{1}{t^2}+ \frac{1}{u^2}\biggr)\biggr]
\end{split}
\end{equation}
\end{widetext}
where the electron mass $m_e$  is neglected for simplicity. This result is consistent with the results in the literature~\cite{Brown79}. The above coupling term can be also written in term of chiral coupling constants as $(c_V^{e})^4+(c_A^{e})^4+6 (c_V^{e})^2 (c_A^{e})^2=8 (c_L^{e})^4+8(c_R^{e})^4$. Then, the integrated cross section is obtained by
\begin{equation} \label{eq:totalsigma}
\sigma_{\text{Born}}^{e^{-}e^{+} \rightarrow Z Z}=\frac
{1}{2!}\int_{t_{\text{min}}}^{t_{\text{max}}} \biggl(\frac{d\sigma}{d t }\biggr)_{\text{Born}}^{e^{-}e^{+} \rightarrow ZZ } d t
\end{equation}
where the lower and upper bounds of the integral are defined as
\begin{equation}
t_{\text{max},\text{min}}=M_{Z}^2-\frac{1}{2} s\bigl( 1 \pm \kappa \bigr).
\end{equation}

\subsection{One-loop EW radiative corrections}
The radiative corrections consist of three parts. The first comes from virtual loop corrections, the second one
is soft photon radiation and the last one corresponds to hard real photon radiation. We now discuss them for $e^{-}e^{+} \rightarrow ZZ$ below.
\subsubsection{Virtual corrections}
\begin{figure*}[htb]
    \begin{center}
\includegraphics[width=0.95\linewidth]{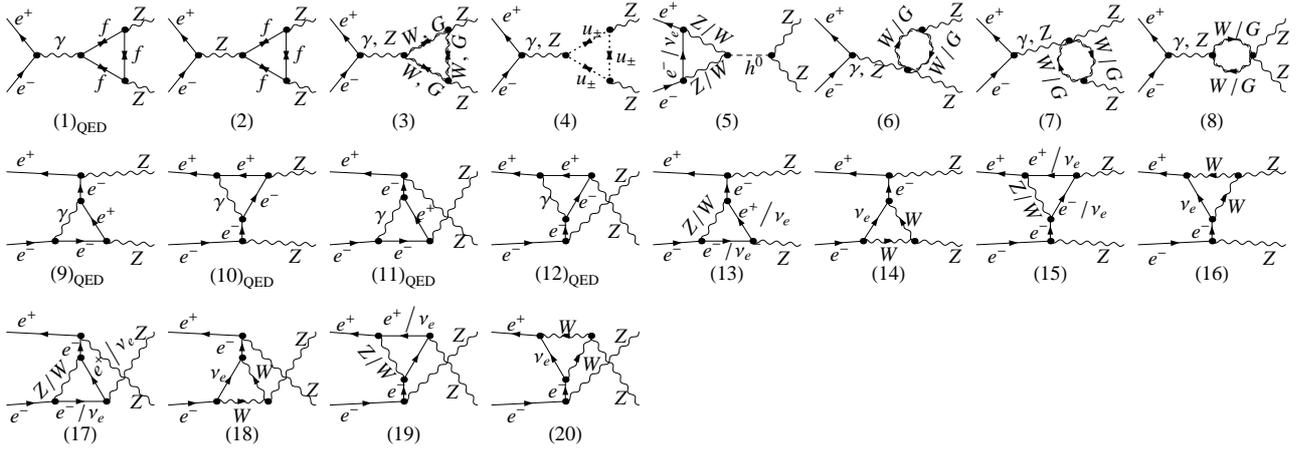}
     \end{center}
     \vspace{-5mm}
\caption{The vertex-correction diagrams contributing to $e^{-}e^{+} \rightarrow ZZ$.}\label{fig:diagvert}
\end{figure*}

\begin{figure}[bh]
    \begin{center}
\includegraphics[width=1\linewidth]{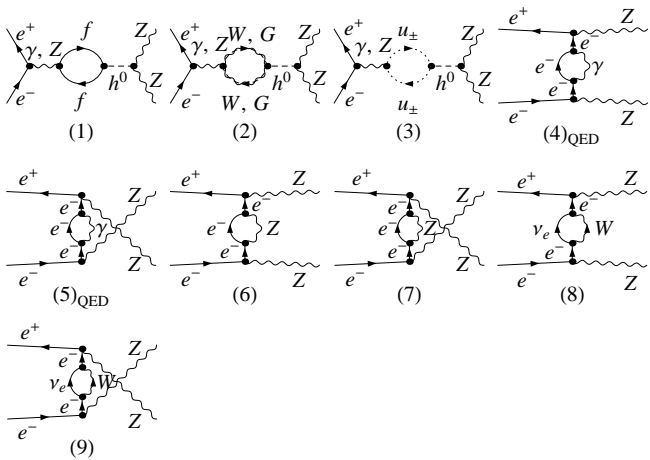}
     \end{center}
     \vspace{-5mm}
\caption{The self-energy diagrams contributing to $e^{-}e^{+} \rightarrow ZZ$.}\label{fig:diagself}
\end{figure}

We consider full EW $\mathcal{O}(\alpha)$ contributions at one-loop level for the process $e^{-}e^{+} \rightarrow ZZ$. The virtual loop corrections for this process come from three types of diagrams: vertex-type, self-energy, and box-type. Here, we do not provide their explicit analytical expressions since they are not particularly illuminating.  Instead, we present a complete list of Feynman diagrams created by the \textsc{FeynArts}. At one-loop order, the process~\eqref{eq:eeZZ} has a total of 161 one-loop diagrams (112 vertex-type+34 self-energy+ 15 box-type) as shown in Figs.~\ref{fig:diagself} to~\ref{fig:diagvert}.
We use the following labels on the internal lines: $f$ stands for all fermions in SM, $G$ is the charged Goldstone boson, and $u_{\pm}$ is the ghosts. The gauge vector bosons ($\gamma$, $Z$, and $W^\pm$) are denoted by wavy lines, and the Higgs and Goldstone bosons are denoted by the dashed lines. In diagrams with two arrows on the same lines of the loop, particles move both clockwise and counterclockwise.

The one-loop diagrams can topologically divided into $s$, $t$, and $u$-channels with the mediator of gauge bosons ($\gamma$, $Z$, $W^\pm$), Higgs boson ($h^0$), and charged Goldstone bosons ($G^\pm$). We can also separate them into QED and Weak corrections. The diagrams of QED virtual corrections are obtained by the possible virtual photon attachment and fermion loop insertion to the Born-level diagrams.

First, in Fig.~\ref{fig:diagvert}, we present the vertex-correction diagrams, which consist of triangle corrections to $t-$channel electron exchange, bubbles, and triangle vertices attached to the initial or final state via an mediator  of $\gamma$, $Z$, and $h^{0}$. They can also be divided into three classes. The first comes from the vertex corrections $Z Z A^{*}/Z^{*}$. Here, the off-shell field is marked by asterisk. These are given with diagrams $(1)$-$(4)$ and $(6)$-$(8)$ in Fig.~\ref{fig:diagvert}. The second is due to the vertex corrections $e e h^{0*}$ and it is shown in diagram $(5)$ in Fig.~\ref{fig:diagvert}. Both types are $s-$channel contributions.  The fermion loop contributions to the vertex of $e e h^{0*}$ are proportional to the electron mass and hence is suppressed. The third type is the vertex corrections $Z e \overline{e}^{*}$ in the $t-$ and $u-$ channels. The $Z e \overline{e}^{*}$ correction is shown in diagrams $(9)-(20)$ of Fig.~\ref{fig:diagvert}.

Second,  in Fig.~\ref{fig:diagself}, we show the self-energy diagrams, which consist of all possible loops of $e^-$ or $\nu_{e^-}$ with gauge bosons $\gamma$, $Z$, or $W^\pm$ on the electron propagator via $t-$ and $u-$channels. Also, there are all possible loops of fermions, $W^\pm$ and $G^\pm$ bosons on the $\gamma h^0$ or $Z h^0$ mixing propagator in $s-$channel. Diagrams (4) and (5) in Fig.~\ref{fig:diagself} are obtained by QED correction, while others are weak corrections.

\begin{figure}[b]
    \begin{center}
\includegraphics[width=1\linewidth]{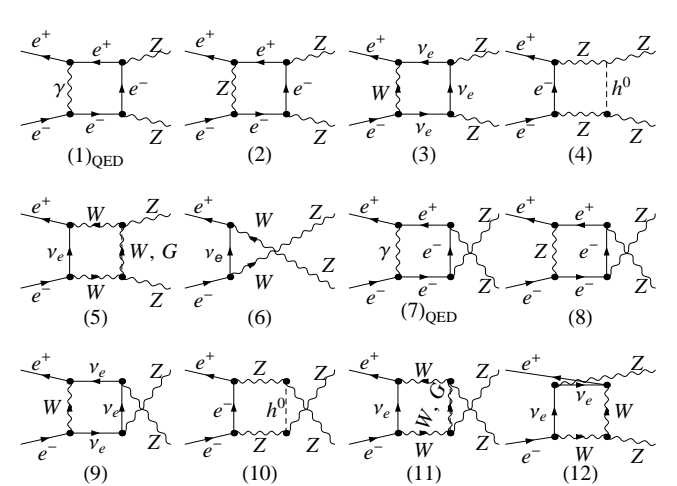}
     \end{center}
          \vspace{-5mm}
\caption{The box-type diagrams contributing to $e^{-}e^{+} \rightarrow ZZ$.}\label{fig:diagbox}
\end{figure}
Finally, Figure~\ref{fig:diagbox} shows the irreducible one-loop diagrams, i.e., the box-type corrections. They consist of all possible box-loops of $e^-$, $\nu_{e^-}$, vector bosons $\gamma$, $Z$, $W^\pm$, Higgs and Goldstone bosons. These are mainly $t$ and $u$-channel contributions. Diagrams (1) and (7) in Fig.~\ref{fig:diagbox} are obtained by QED correction, while others are weak corrections. Diagrams (2), (4),(8) and (10) are due to neutral current correction and other diagrams are due to charged current correction.

The total amplitude for the virtual contributions is defined as the summation of triangle-type, self-energy and box-type  corrections:
\begin{equation}\label{eq:totalM}
{\cal \delta M}_{\text{virt}}= {\cal M}_{\triangle}+{\cal M}_{\bigcirc} + {\cal M}_{\Box}.
\end{equation}
The differential cross section of the virtual corrections is calculated by
\begin{eqnarray} \label{eq:dsigmavirt}
\begin{split}
\biggl(\frac{d\sigma}{d\Omega }\biggr)_{\text{virt}}^{e^{-}e^{+} \rightarrow ZZ } = &\sum_{\sigma_{1,2},\lambda_{1,2}}\frac{1}{4} (1+2 \sigma_1 P_1)(1+2 \sigma_2 P_2)\\
& \times \frac
{\kappa}{64 \pi^2 s}  2 \text{Re}\bigl[{\cal M}_{\text{Born}}^*{\cal \delta M}_{\text{virt}}\bigr],
\end{split}
\end{eqnarray}
where $|{\cal \delta M}_{\text{virt}}|^2$ is not included because its contribution is so small that it can be neglected. The one-loop Feynman diagrams, which form the virtual $\mathcal{O}(\alpha)$ corrections ${\cal \delta M}_{\text{virt}}$, have been calculated in 't Hooft-Feynman gauge. We use the on-shell (OS) renormalization scheme (see, Ref.~\cite{Denner93}) to fix all the renormalization constants.

The virtual contributions have ultraviolet (UV) and infrared (IR) divergences. These divergences can be regularized by extending
the dimensions of spinor and spacetime manifolds to $D=4-2\epsilon$~\cite{Hooft72} and adding a photon mass parameter, respectively.
\begin{figure}[hbt]
    \begin{center}
\includegraphics[width=1\linewidth]{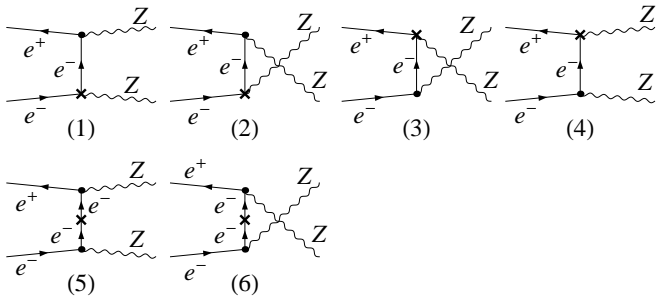}
     \end{center}
     \vspace{-5mm}
\caption{The counterterm diagrams.}\label{fig:counter}
\end{figure}
We adopt all Feynman rules of the counterterms (shown in Fig.~\ref{fig:counter}) and of the renormalization conditions from Ref.~\cite{Denner93}. The redefinition of parameters and fields is carried out in the OS scheme. This turns the Lagrangian into a bare and counterterms.
After applying the renormalization procedure, a UV-finite virtual contribution is achieved. Although, the soft IR singularity due to virtual photonic loop corrections still exists. It is regulazed by giving the photon a fictitious mass, $m_\gamma$. The virtual cross section is independent of the UV regularization parameter $C_{UV} =1/\epsilon-\gamma_E + \log (4 \pi)$, but still a function of the IR regularization parameter $m_\gamma$. From the Kinoshita-Lee-Nauenberg theorem~\cite{Kinoshita62,Lee64}\footnote{This was also shown perturbatively in QED by Schwinger~\cite{Schwinger49}.}, it is cancelled in the limit $m_\gamma \rightarrow 0$ by adding the real photon corrections. We have checked numerically that our results do not depend on $m_\gamma$. After adding the virtual and real corrections, the results are still collinear singular. This singularity comes from the initial state radiation part. We use the phase space slicing method\footnote{We have developed the necessary code for evaluating of collinear contributions and implemented in \textsc{FormCalc}.}~\cite{PSS1,PSS2,PSS3,PSS4} to handle the collinear singularities in the photon radiation off initial state.

\subsubsection{Real corrections}
Real photon emission gives rise to the kinematically
different reaction from $e^{+}e^{-} \rightarrow ZZ$, presented by
\begin{equation} \label{eq:eeZZgamma}
e^{+}(p_1,\sigma_1)e^{-}(p_2,\sigma_2)\rightarrow Z(k_1,\lambda_1) Z(k_2,\lambda_2) \gamma(k_3,\lambda_3),
\end{equation}
where $k_3$ and $\lambda_3$ are four-momenta and helicity of the radiated photon.
\begin{figure}[ht]
    \begin{center}
\includegraphics[width=1\linewidth]{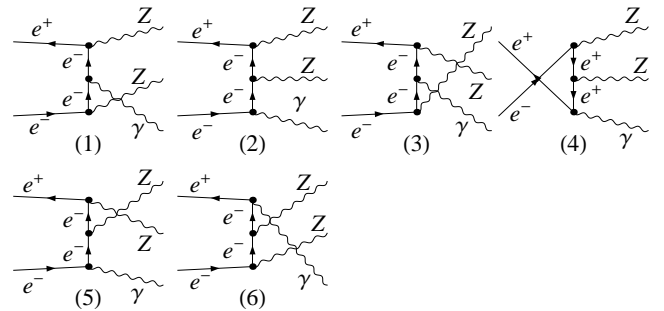}
     \end{center}
     \vspace{-5mm}
\caption{The Feynman diagrams for the real photon radiation.}\label{fig:radiation}
\end{figure}
We present the bremsstrahlung Feynman diagrams in Fig.~\ref{fig:radiation}, obtained from the Born diagrams by adding a photon emitted from a lepton line. The lowest-order of real-photon emission provides an $\mathcal{O}(\alpha)$ correction to $e^{+}e^{-} \rightarrow ZZ$.

The differential cross section of $e^{+}e^{-} \rightarrow ZZ\gamma$ reads
\begin{equation} \label{eq:difsigma}
\begin{split}
\biggl(\frac{d\sigma}{d\Omega_3 }\biggr)_{\text{real}}^{e^{+}e^{-} \rightarrow ZZ\gamma} = &\frac
{1}{2 s} \sum_{\sigma_{1,2},\lambda_{1,2,3}}\frac{1}{4} (1+2 P_1 \sigma_1)(1+2P_2 \sigma_2)\\
& \times |{\cal M}_{e^{+}e^{-} \rightarrow ZZ\gamma}^{\sigma_1,\sigma_2,\lambda_1,\lambda_2,\lambda_3}|^2
\end{split}
\end{equation}
with the 3-particle phase-space integral
\begin{equation} \label{eq:domega3}
\begin{split}
\int d\Omega_3 = \prod_{i=1}^{3} \int \frac{d^3 \vec{k}_i}{(2 \pi)^3 2k_i^0 }(2 \pi)^4
\delta\biggl(p_1+p_2-\sum_{j=1}^3 k_j \biggr).
\end{split}
\end{equation}

According to the radiated photon energy $E_\gamma=\sqrt{|\overrightarrow{k_3}|^2+m_\gamma^2}$, the bremsstrahlung phase space can be separated into two regions: soft and hard. Hence, the real photon radiation correction can be organized as
\begin{equation} \label{eq:dsigmaSB}
d\sigma_{\text{real}}^{e^{-}e^{+}\rightarrow ZZ\gamma}=d\sigma_{\text{soft}}(\Delta_s)+d\sigma_{\text{hard}}(\Delta_s)
\end{equation}
where $\Delta_s=\Delta E_\gamma/(\sqrt{s}/2)$ is the soft cutoff energy parameter. The bremsstrahlung photon is named soft when $E_\gamma<\Delta E_\gamma=\Delta_s \sqrt{s}/2 $, whereas it is hard if $E_\gamma>\Delta E_\gamma$. The soft photon correction is given by approximation~\cite{Hooft79,Denner93}
\begin{equation} \label{eq:dsoft}
\begin{split}
d\sigma_{\text{soft}}=-d\sigma_{\text{Born}}\frac{\alpha Q^2_\ell}{2\pi^2} \int_{|\overrightarrow{k_3}|\leq \Delta E_\gamma} \frac{d^3k_3}{2E_\gamma} \biggl[\frac{k_1}{k_1\cdot k_3}-\frac{k_2}{k_2\cdot k_3}\biggl]^2
\end{split}
\end{equation}
where $\Delta E_\gamma$ satisfies $E_\gamma\leq\Delta E_\gamma \ll\sqrt{s}$. Integrating the soft photon phase space in the center-of-mass system gives
\begin{equation} \label{eq:dsoft0}
\begin{split}
d\sigma_{\text{soft}}=\delta_{\text{soft}}d\sigma_{\text{Born}}
\end{split}
\end{equation}
with
\begin{equation} \label{eq:dsoft1}
\begin{split}
\delta_{\text{soft}}=&-\frac{\alpha}{\pi}\biggl[2 \ln\biggl( \frac{2\Delta E_\gamma}{m_\gamma}\biggr) \biggl(1+\ln\biggl( \frac{m^2_e}{s}\biggr)\biggr)\\
&+ \frac{1}{2}\ln^2\biggl( \frac{m^2_e}{s}\biggr)+\ln\biggl( \frac{m^2_e}{s}\biggr)+ \frac{\pi^2}{3}
\biggr].
\end{split}
\end{equation}

Both the soft and hard photon corrections depend on the soft cutoff parameter $\Delta_s$, but the real correction is independent of this parameter. We note that the hard photon emission must be taken account into to remove this dependency. Furthermore, adding the virtual and soft corrections removes the $m_\gamma$ dependency.

In addition to the divergences mentioned above, there also appear mass singularities as a consequence of the collinear photon emission off massless particles, so-called collinear divergences. However, the smallness of the electron mass induces the quasi-collinear IR divergences from the photon radiation off the electron/positron. To deal with this, we apply the phase space slicing method. According to this method, the hard bremsstrahlung phase space can be divided into collinear and finite regions:
\begin{equation} \label{eq:dsigmahard}
d\sigma_{\text{hard}}^{e^{-}e^{+}\rightarrow ZZ\gamma}(\Delta_s)=d\sigma_{\text{coll}}(\Delta_s,\Delta_c)+d\sigma_{\text{fin}}(\Delta_s,\Delta_c)
\end{equation}
where $\Delta_c$ denotes the angular cut-off parameter. In the hard collinear region $(E_\gamma\geq\Delta E_\gamma, \cos \theta_{e\gamma}>1-\Delta_c)$, the integrand is numerically unstable, whereas in the hard finite region $(E_\gamma\geq\Delta E_\gamma, \cos \theta_{e\gamma}\leq 1-\Delta_c)$, it is  finite (numerically stable). Here, $\cos \theta_{e\gamma}$ is the cosines of angle between the electron/positron and bremsstrahlung photon. In the stable regions the integration is carried out numerically, while it is semi-analytically calculated in the unstable regions ~\cite{PSS5} by using approximation:
\begin{equation} \label{eq:dcoll}
\begin{split}
d\sigma_{\text{coll}}=&\sum_{i=1}^2 \frac{\alpha}{2\pi}  Q^2_i\int^{1-\Delta_s}_0 dx~d\sigma_{\text{Born}}(\sqrt{x s}) \\
&\times\biggl[\frac{1+x^2}{1-x} \ln \biggl(\frac{s \Delta_c}{2 m_i^2}\biggr)-\frac{2x}{1-x} \biggl]
\end{split}
\end{equation}
where the approximation $\Delta_c\gg 2 m_e^2/s$  has been taken.
To avoid over-counting in the soft energy region, the integration over all possible factors $x$ is constrained by the soft cut-off parameter $\Delta_s$.

\subsubsection{Classification of full corrections}
As a result, the IR and UV finite EW corrections consist of four parts:
\begin{equation} \label{eq:dsigmaNLO}
\begin{split}
d\sigma_{\text{NLO}}^{e^{+}e^{-}\rightarrow ZZ }&=d\sigma_{\text{virt}}(m_\gamma)+d\sigma_{\text{soft}}(m_\gamma,\Delta_s)\\
&+d\sigma_{\text{coll}}(\Delta_s,\Delta_c)+d\sigma_{\text{fin}}(\Delta_s,\Delta_c).
\end{split}
\end{equation}
where they are from virtual corrections, soft photon corrections, collinear corrections and finite hard photon corrections, respectively.

\begin{figure*}[hbt]
\centering
\includegraphics[width=0.39\textwidth]{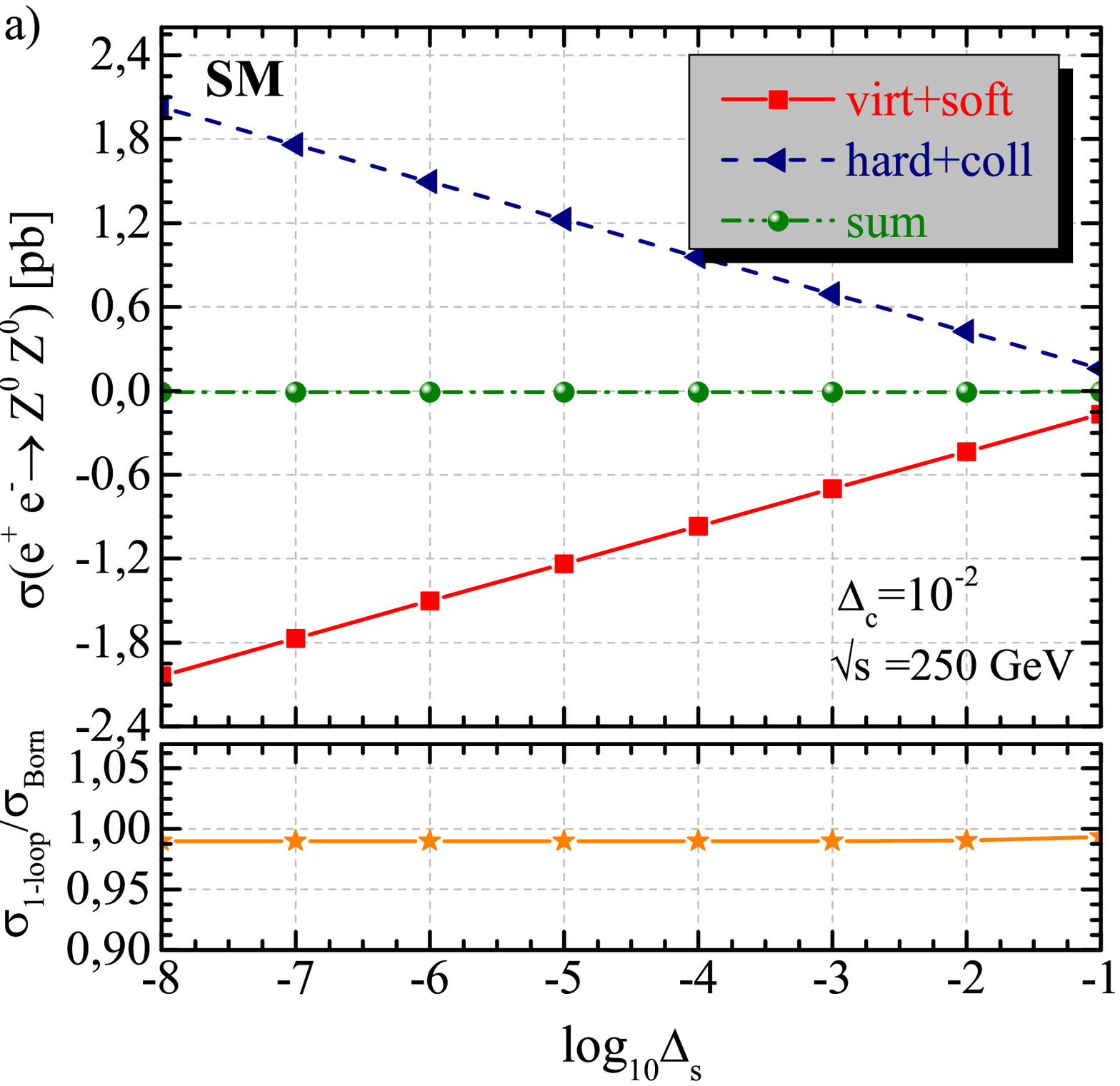}
\vspace{1em}
\begin{minipage}[b]{0.4\textwidth}
\centering
\begin{tabular}[b]{lr}
\toprule
$\Delta_s$ & $\sigma_{\text{1-loop}}/\sigma_{\text{Born}}$ \\
\midrule
$10^{-1}$ & $0.9933$ \\
$10^{-2}$ & $0.9904$ \\
$10^{-3}$ & $0.9899$ \\
$10^{-4}$ & $0.9899$ \\
$10^{-5}$ & $0.9899$ \\
$10^{-6}$ & $0.9898$ \\
$10^{-7}$ & $0.9897$ \\
$10^{-8}$ & $0.9897$ \\
\bottomrule
\end{tabular}
\vspace{2em}
\end{minipage}
\centering
\includegraphics[width=0.39\textwidth]{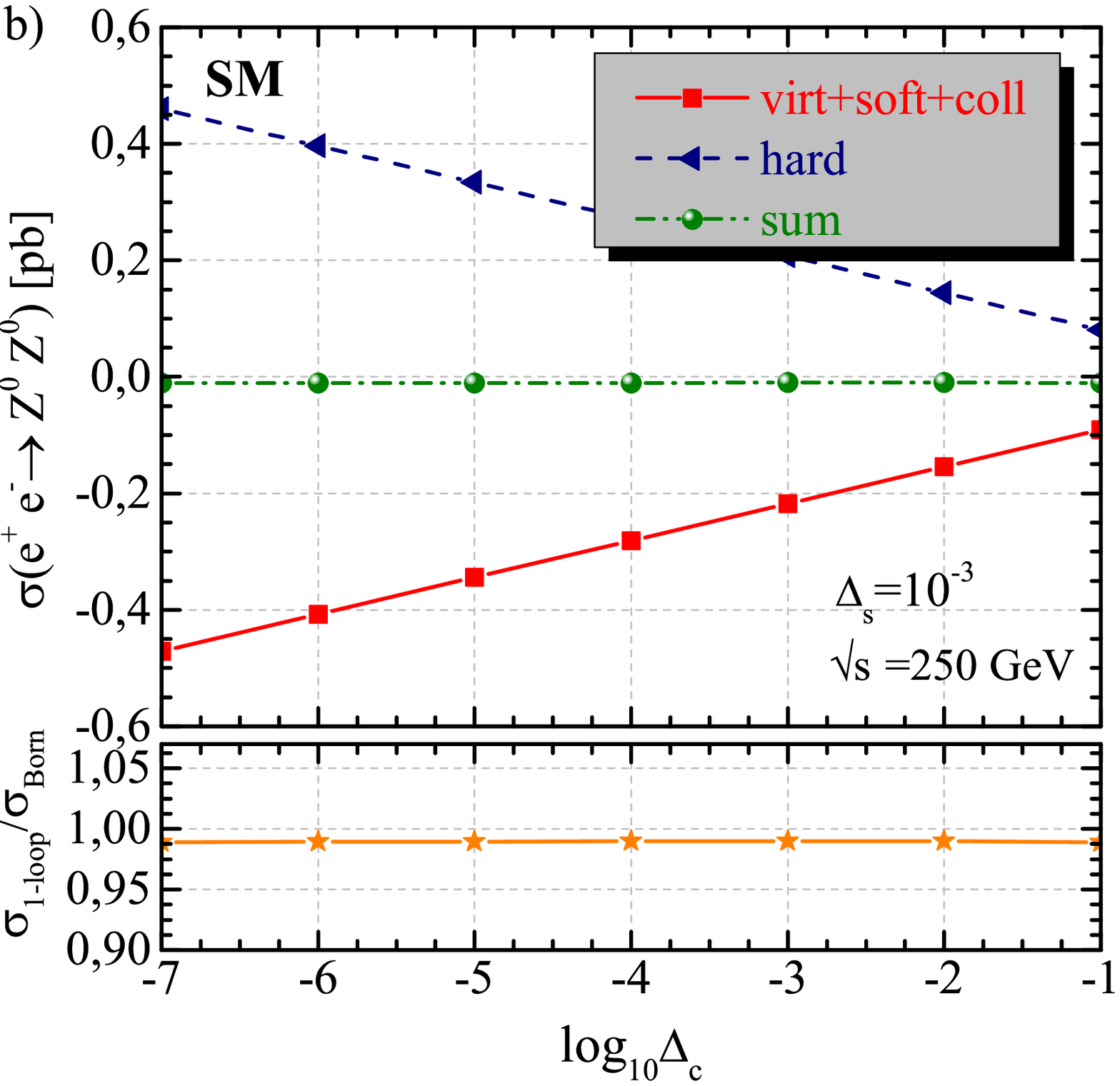}
\begin{minipage}[b]{0.4\textwidth}
\centering
\begin{tabular}[b]{lr}
\toprule
$\Delta_c$ & $\sigma_{\text{1-loop}}/\sigma_{\text{Born}}$ \\
\midrule
$10^{ -1}$ & $0.9891$ \\
$10^{-2}$ & $0.9899$ \\
$10^{-3}$ & $0.9899$ \\
$10^{-4}$ & $0.9897$ \\
$10^{-5}$ & $0.9894$ \\
$10^{-6}$ & $0.9892$ \\
$10^{-7}$ & $0.9889$ \\
\bottomrule
\end{tabular}
\vspace{2em}
\end{minipage}
\vspace{-4mm}
\caption{\label{fig:PSS}
  Phase space slicing method.  The virtual, soft and hard photon radiation corrections as a function of soft cut-off parameter $\Delta_s$ (upper plot) and angular cut-off parameter $\Delta_c$ (lower plot) at $\sqrt{s} = 250\gev$.
}
\end{figure*}
We have numerically checked that our results are independent of $m_\gamma$, soft cut-off parameter $\Delta_s$ and angular cut-off parameter $\Delta_c$. The virtual, soft photon and hard photon emission corrections are shown as a function of both $\Delta_s$ and $\Delta_c$ at $\sqrt{s}=250\gev$ in Fig.~\ref{fig:PSS}. It is obvious that the virtual plus soft and the hard photon emission corrections significantly depend on $\Delta_s$ and $\Delta_c$, whereas the total corrections do not depend on them. Moreover, the relative correction $\sigma_{\text{1-loop}}/\sigma_{\text{Born}}$ is independent of $\Delta_s$ and $\Delta_c$ over several orders of magnitude. In order to more clearly show the cut-off independence, we have also given the values by tables next to Fig.~\ref{fig:PSS}. Finally, our results are also stable over nine digits when varying $m_\gamma$ from $10^{-20}$ GeV to the default value of $1$ GeV. Our numerical results below have been obtained for $\Delta_s=10^{-3}$, $\Delta_c=10^{-2}$ and the range of scattering angles of the final particles $|\cos\theta| < 0.99$.

In order to discuss the origin of the large correction, we also consider the pure QED corrections $\delta_{\text{QED}}$ as individual in addition to the full EW corrections $\delta_{\text{Total}}$. The QED corrections consist of virtual-photon exchange, the corresponding counterterms and real-photon emission. Hence, we can express the QED relative correction as
\begin{equation} \label{eq:dqed}
\delta_{\text{QED}}=\delta_{\text{virt, QED}}+\delta_{\text{real}}.
\end{equation}
The QED-like diagrams consist of only $A$ and $f$ fields as virtual lines. In this study, the one-loop QED contributions come from the sum of the real-photonic corrections and contributions of diagrams $(4)$ and $(5)$ of Fig.~\ref{fig:diagself}, $(1)$ and $(7)$ of Fig.~\ref{fig:diagbox}, and $(1)$ and $(9)-(12)$ of Fig.~\ref{fig:diagvert}.

The remaining corrections (non-QED) can be called as Weak corrections $\delta_{\text{Weak}}$, which include the massive gauge bosons $Z^0$ and $W^\pm$. The genuine weak relative correction can be obtained from
\begin{equation} \label{eq:dweak}
\delta_{\text{Weak}}=\delta_{\text{Total}}-\delta_{\text{QED}}.
\end{equation}
In the above definition of $\delta_{\text{Weak}}$, the Weak corrections are obtained by subtracting the pure QED corrections from the full EW radiative corrections. Thus, we can write the full $\mathcal{O}(\alpha)$ EW relative correction as
\begin{equation} \label{eq:dtotal}
\delta_{\text{Total}}=\delta_{\text{QED}}+\delta_{\text{Weak}}.
\end{equation}

Overall, we can factorize the full EW corrected cross section into the Born cross section and the relative corrections. Therefore, the one-loop cross section $\sigma_{\text{1-loop}}$ becomes
\begin{equation} \label{eq:sigma}
\begin{split}
\sigma_{\text{1-loop}}&=\sigma_{\text{Born}}(1+\delta_{\text{Total}})\\
&=\sigma_{\text{Born}}(1+ \delta_{\text{QED}}+\delta_{\text{Weak}}),
\end{split}
\end{equation}
leading to
\begin{equation} \label{eq:dtotal}
\begin{split}
\delta_{\text{X}}&=\frac{\sigma^{\text{X}}_{\text{1-loop}}-\sigma_{\text{Born}}}{\sigma_{\text{Born}}}
\end{split}
\end{equation}
where $\text{X}$ can be ``QED'', ``Weak'' and ``Total''.

\section{Beam Polarization} \label{sec:pol}
Polarization effects are important in $e^+e^-$ colliders and can be used to provide significant advantages. This effects could benefit to searches for new physics with small deviation from the SM predictions in two ways.
First, properly chosen combinations of beam polarization can strengthen the signal and suppress the background in many cases. Second, it is possible to establish smart observables that contain beam polarization information. With the above motives, we analyse the effect of beam polarizations on the production rate of $e^+e^- \to ZZ$.

The polarizations with a sign for $e^-$ and $e^+$ beams are given by~\cite{Omori99}, respectively
\begin{equation}  \label{eq:Pe}
\begin{split}
P_{ e^-}&=\frac{(n_{e^-_R}-n_{e^-_L})}{(n_{e^-_R}+n_{e^-_L})},\,\,\
P_{ e^+}=\frac{(n_{e^+_R}-n_{e^+_L})}{(n_{e^+_R}+n_{e^+_L})}
\end{split}
\end{equation}
where $n_{e^\pm_L}$ and $n_{e^\pm_R}$ are the number of the left- and right-handed  $e^\pm$'s ($e^\pm_L$ and $e^\pm_R$) in the beam, respectively. Here, $P_{ e^-}$/$P_{ e^+}$ equal to $+1$ (-1) for the $100\%$ right-(left)-handed polarized $e^-/e^+$-beams.

\begin{figure}[!hbt]
    \begin{center}
\includegraphics[width=1\linewidth]{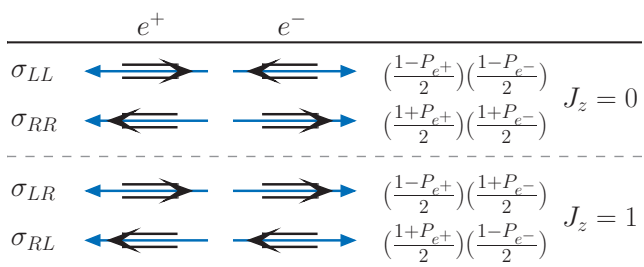}
     \end{center}
     \vspace{-6mm}
\caption{The longitudinal spin configurations in electron-positron collisions. The thick arrow denotes the direction of motion of the particle
and the double arrow its spin direction.}
\label{fig:spinconf}
\end{figure}

If a normalization $n_{e^\pm_L}+n_{e^\pm_R}=1$ is applied, the normalized number of $e^-_R$'s and $e^-_L$'s can be obtained as
\begin{equation}  \label{eq:nPe}
\begin{split}
n_{e^-_R}=\frac{1+P_{ e^-}}{2},\,\,\ n_{e^-_L}=\frac{1-P_{ e^-}}{2}.
\end{split}
\end{equation}
Consequently, the cross section for any beam polarizations can be defined by~\cite{Hikasa86,Omori99,Moortgat08}
\begin{equation}  \label{eq:polsigma}
\begin{split}
\sigma^{P_{ e^+}P_{ e^-}}=&\frac{1}{4}\big[( 1-P_{ e^+})(1-P_{ e^-})\sigma_{LL}\\
&+  ( 1+P_{ e^+})(1+P_{ e^-})\sigma_{RR} \\
&+( 1-P_{ e^+})(1+P_{ e^-})\sigma_{LR}\\
&+( 1+P_{ e^+})(1-P_{ e^-})\sigma_{RL}\big],
\end{split}
\end{equation}
where $\sigma_{LL}$, $\sigma_{RR}$, $\sigma_{LR}$, and $\sigma_{RL}$ indicate the cross sections with completely polarized beams of the four possible cases. Namely, $\text{RL}, \text{LR}, \text{RR}$ and $\text{LL}$ stand for $(P_{e^+}, P_{e^-})= (+1, -1), (-1,+1), (+1,+1), (-1,-1)$, respectively. Figure~\ref{fig:spinconf} shows these spin configurations, the corresponding fractions (the fourth column) and the total spin projections onto the $e^+e^-$ direction (the last column). This figure is adapted from Ref.~\cite{Moortgat08}.

Now we give an expression of the left-right asymmetry $A_{LR}$, which has several advantages such as it is independent of detector efficiency asymmetries and its measurement has negligible systematic error.
It is defined by
\begin{eqnarray}
A_{LR}=\frac{\sigma(P_{ e^+}=+1, P_{ e^-}=-1)-\sigma(P_{ e^+}=-1, P_{ e^-}=+1)}{\sigma(P_{ e^+}=+1, P_{ e^-}=-1)+\sigma(P_{ e^+}=-1, P_{ e^-}=+1)} \nonumber
\end{eqnarray}
or equivalently
\begin{equation}  \label{eq:ALR}
\begin{split}
A_{LR}&=\frac{\sigma_{RL}-\sigma_{LR}}{\sigma_{LR}+\sigma_{RL}}
\end{split}
\end{equation}
for a given process.
For the process $e^+e^- \to ZZ$ considered in this study, it is obtained by
\begin{equation}  \label{eq:ALRBORN}
\begin{split}
A_{LR}^{\text{Born}}=\frac{(c_L^e)^4-(c_R^e)^4}{(c_L^e)^4+(c_R^e)^4}=0.40979
\end{split}
\end{equation}
at the Born-level.

If $e^+e^-$ is annihilated into a vector particle, only $\sigma_{LR}$ and $\sigma_{RL}$ ($J = 1$ configurations) have a non-zero contribution. In this study, the $\sigma_{LL}$ and $\sigma_{RR}$ have the tiny values due to small electron mass effects, so they can be neglected. In this case, we can also rewrite the left-right asymmetry as~\cite{Moortgat08}
\begin{equation}  \label{eq:ALRobs}
\begin{split}
A_{LR}=\frac{1}{P_{eff}}A_{LR}^{\text{obs}}=\frac{1}{P_{eff}}\biggl(\frac{\sigma_{-+}-\sigma_{+-}}{\sigma_{-+}+\sigma_{+-}}\biggr)
\end{split}
\end{equation}
in terms of the effective polarization $P_{eff}$ and the measured left-right asymmetry $A_{LR}^{\text{obs}}$
\begin{equation}  \label{eq:Peff}
P_{eff}=\frac{(P_{ e^-}-P_{ e^+})}{(1-P_{ e^-}P_{ e^+})},\,\,A_{LR}^{\text{obs}}=\frac{\sigma_{-+}-\sigma_{+-}}{\sigma_{-+}+\sigma_{+-}},
\end{equation}
where the corresponding cross-sections are given by
\begin{equation}  \label{eq:sigmapm}
\begin{split}
\sigma_{-+}=&\frac{1}{4}\biggl[\bigl(1+|P_{ e^+}||P_{ e^-}|\bigr)\bigl(\sigma_{RL}+\sigma_{LR}\bigr)\\
&+\bigl(|P_{ e^+}|+|P_{ e^-}|\bigr)\bigl(\sigma_{RL}-\sigma_{LR}\bigr)\biggr]\\
\sigma_{+-}=&\frac{1}{4}\biggl[\bigl(1+|P_{ e^+}||P_{ e^-}|\bigr)\bigl(\sigma_{RL}+\sigma_{LR}\bigr)\\
&-\bigl(|P_{ e^+}|+|P_{ e^-}|\bigr)\bigl(\sigma_{RL}-\sigma_{LR}\bigr)\biggr].
\end{split}
\end{equation}
Another parameter, the effective luminosity $\mathcal{L}_{eff}$ is given by
\begin{equation}  \label{eq:Leff}
\frac{\mathcal{L}_{eff}}{\mathcal{L}}=\frac{1}{2}(1-P_{ e^-}P_{ e^+})
\end{equation}
which basically reflects the fraction of particles that are interacting.
\section{Parameter Settings} \label{sec:inputs}
A set of input parameters must be specified with their corresponding numeric values, in order to provide consistent higher-order predictions in the SM. We set the input parameters as follows, see also~\cite{PDG20}:
\begin{itemize}
  \item Mass parameters:

\begin{equation*}\label{eq:SMpar}
\begin{tabular}{ll}
$m_e=0.510998928\mev$,&$m_u=73.56\mev$  \\
$m_\mu=105.6583715\mev$,&$m_d=73.56\mev$, \\
$m_\tau=1.77682\gev$, & $m_s=95\mev$, \\
$M_W= 80.385\gev$, &$m_c=1.275\gev$,\\
$M_Z= 91.1876\gev$, & $m_b=4.66\gev$, \\
$M_h= 125\gev$, &$m_t=173.21\gev$.\\
\end{tabular}
\end{equation*}
Here, the $u$- and $d$-quark masses are calculated as a effective parameters and are specially given for the $M_Z$-mass scale via the hadronic contributions 
\begin{align} \label{eq:total_cross}
\begin{split}
\Delta\alpha_{\text{had}}^{(5)}(M_Z) &=
      \frac{\alpha}{\pi}\sum_{f = u,c,d,s,b}
      Q_f^2 \Bigl(\ln\frac{M_Z^2}{m_f^2} - \frac 53\Bigr) \\
      &\approx 0.027547\,.
\end{split}
\end{align}
According to ~\cite{PDG20}, $s$-quark mass $m_s$ is an estimate of a so-called "current quark mass" in the $\overline{MS}$ scheme at scale $\mu \approx 2\gev$.  $m_c \equiv m_c(m_c)$ is ``running'' mass in the $\overline{MS}$ scheme  and $m_b$ is the $\Upsilon(1S)$ bottom quark mass.

  \item The fine structure constant:
  \begin{equation}
   \alpha(0)= 1/137.03599907
   \end{equation}
  \item The Fermi constant:
  \begin{equation}
  G_F = 1.1663787(6)\times10^{-5}\gev^{-2}
  \end{equation}
  which is conventionally defined via the muon lifetime.
\end{itemize}
 The renormalization scale $\mu_R$ is fixed to center-of-mass energy,
$\sqrt{s}$.

On the other hand, it is important to specify the electromagnetic coupling $\alpha = e^2/(4\pi)$ for the EW $\mathcal{O}(\alpha)$ corrections. For an obvious choice of $\alpha$, there are two different methods as follows: the fine-structure constant $\alpha(0)$ in the Thompson limit ($\alpha(0)$ scheme) and the running electromagnetic coupling $\alpha(Q^2)$ at any energy scale $Q$. It is possible to use the value of $\alpha(M^2_Z)\approx 1/129$ ($\alpha(M^2_Z)$ scheme), which is calculated by analyzing the experimental ratio $R=\sigma(e^-e^+\rightarrow \text{hadrons})/\sigma(e^-e^+\rightarrow \mu^-\mu^+)$~\cite{Burkhardt95,Eidelman95}. Another choice is a $G_\mu$ scheme given by
\begin{equation}
\alpha(G_\mu) =\frac{\sqrt{2} G_\mu M^2_W}{\pi} \biggl(1-\frac{M^2_W}{M^2_Z}\biggr).
\end{equation}
An effective value of $\alpha$ in this scheme is obtained as $\alpha(G_\mu) \approx 1/132$ from the Fermi constant $G_\mu$.

The $G_\mu$ scheme provides the possibility to absorb some important universal corrections associated with the renormalization of the weak mixing angle into leading order contributions. At next-to-leading order (NLO), the $\alpha(0)$ and $G_\mu$ schemes are related by
\begin{equation}\label{eq:alphNLO}
\alpha(G_\mu) = \alpha(0)\bigl[1 + \Delta r^{(1)}\bigr] + \mathcal{O}(\alpha^3) 
\end{equation}
where $\Delta r^{(1)}$ denotes the EW correction to muon decay at NLO~\cite{Sirlin80,Denner93}. In Sirlin's relation, the resummation is achieved by the replacement  $(1+\Delta r^{(1)}) \rightarrow \frac{1}{(1-\Delta r^{(1)})}$ in Eq.~\eqref{eq:alphNLO}. One then obtains a much closer agreement between the two schemes. This fact also reveals that the $G_\mu$-scheme provides more accurate results above the $M_Z$-mass scale unless one applies the leading log resummation in the strict NLO $\alpha$-scheme.

Actually, the suitability of the scheme directly relates to the nature of the considered process. In all cases, a common coupling factor $\alpha^n$ should be used in full gauge-invariant sub-groups, otherwise significant consistency relations disappear~\cite{Denner20}.

In this study, we present the results obtained by both $\alpha(0)$ scheme and $G_\mu$ scheme, and discuss the difference due to these choices.

\section{Numerical Results And Discussions} \label{sec:results}
We examine the Z boson pair production in electron-positron collision, by taking into account a full one-loop EW $\mathcal{O}(\alpha)$ corrections, including soft and hard QED radiation. We provide the center-of-mass energy dependence of  Born and one-loop cross sections. In order to discuss the origin of the large correction, we also present the relative QED and Weak corrections (as defined in Eq.~\eqref{eq:dtotal}) as a function of the center-of-mass energy. We include the spin polarization
effects of the initial electron and positron beams in the total cross sections. We discuss how the considered process is affected by beam polarization. The left-right asymmetry  and angular distributions are also presented.

First of all, for the numerical verification of Born-level and hard photon bremsstrahlung calculations, we use three different tools \textsc{FeynArts}$\&$\textsc{FormCalc}, \textsc{CalcHEP} and \textsc{Whizard}. In Table~\Ref{table:toolscomp}, we give the results obtained by them for Born level process $e^- e^+ \to ZZ$ and hard process $e^- e^+ \to ZZ \gamma$ at $\sqrt{s}=250$, $500$, and $1000\gev$.
\begin{table}[hbt]
\caption{The Born and hard photon bremsstrahlung cross sections obtained by \textsc{FeynArts}$\&$\textsc{FormCalc} (FA$\&$FC), \textsc{Whizard} and \textsc{CalcHEP}.}\label{table:toolscomp}
\centering
\begin{ruledtabular}
\begin{tabular}{llll}
$~~~\sqrt{s}$ & $250\gev$ & $500\gev$ & $1000\gev$\\
\hline
&\multicolumn{3}{c}{$\sigma^{\text{Born}} (e^- e^+ \to Z^0 Z^0)$ [fb]}\\
\cellcolor[gray]{0.7}FA$\&$FC & 991.500(2)  & 315.463(9) &79.592(8) \\
\cellcolor[gray]{0.8}\textsc{Whizard}  &991.501(5) & 315.464(3) &79.595(1)\\
\cellcolor[gray]{0.9}\textsc{CalcHEP}  &991.50      &315.46     &79.593  \\
\hline
&\multicolumn{3}{c}{$\sigma^{\text{hard}-\gamma} (e^- e^+ \to Z^0 Z^0 \gamma)$ [fb]}\\
\cellcolor[gray]{0.7}FA$\&$FC 	   &144.443(4) & 52.751(9) &14.852(4) \\
\cellcolor[gray]{0.8}\textsc{Whizard}  &144.457(1) & 52.739(5) &14.851(3)\\
\cellcolor[gray]{0.9}\textsc{CalcHEP}  &144.46    & 52.758    &14.859  \\
\end{tabular}
\end{ruledtabular}
\end{table}
It is found that Born-level results are an excellent agreement. The hard photon bremsstrahlung results are in good agreement up to five digits.

\begin{figure}[tb]
    \begin{center}
\includegraphics[scale=0.43]{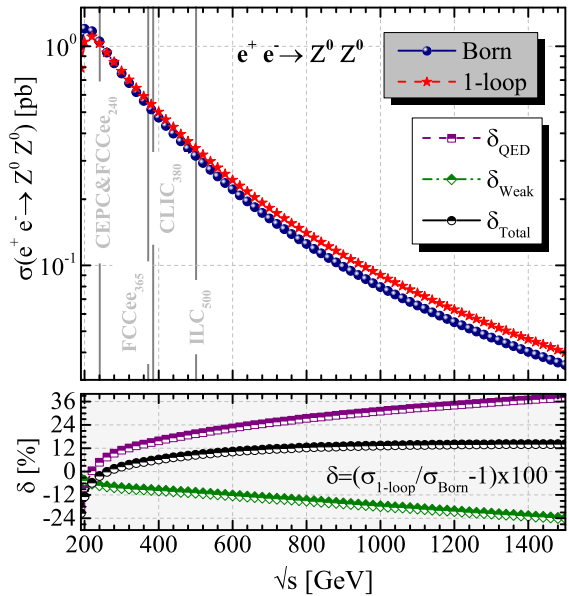}
     \end{center}
     \vspace{-6 mm}
\caption{Born and one-loop cross sections as a function of $\sqrt{s}$. The relative corrections in a percentage is also shown at the bottom panel. The vertical solid lines show the proposed energies for various future colliders.}
\label{fig:cs}
\end{figure}

In Fig.~\ref{fig:cs}, we plot the Born and one-loop cross sections, and the relative corrections as a function of center-of-mass energy. We also show the proposed energies of future colliders with the vertical solid lines. The center of mass energy ranges from 190 GeV to 1.5 TeV in the steps of 10 GeV. 
Since the colliding energy $\sqrt{s}$ starts near the threshold of $2m_Z$, the Born and one-loop cross sections increase quickly with the opening of the phase space, reach a maximum value and then fall off rapidly with the increment of $\sqrt{s}$. This is also the expected behavior. The one-loop cross section reaches a maximum of 1109.93 fm with $\delta_\text{Total}=-8.45\%$ at $\sqrt{s}=210\gev$ and then decreases to 40.04 fb with $\delta_{\text{Total}}=+14.17\%$ at $\sqrt{s}=1.5$ TeV. The pure QED corrections make a positive contribution and increase from $-8.1\%$ to $+31.15\%$ when $\sqrt{s}$ goes from 200 GeV to 1.0 TeV. On the other hand, the weak corrections make a negative contribution and its relative correction decreases from $-4.63\%$ to  $-17.69\%$. The QED and weak corrections partially compensate each other, providing relative corrections of around $-12.73\%$ at the first point and $+13.46\%$ at 1.0 TeV. The EW radiative corrections significantly increase with the $\sqrt{s}$. This is due to the presence already at the one-loop order, of large double and single logarithm terms behaving like
$(\alpha/\pi) \ln^2 s$ and $(\alpha/\pi) \ln s$.

\begin{figure}[b]
    \begin{center}
\includegraphics[scale=0.43]{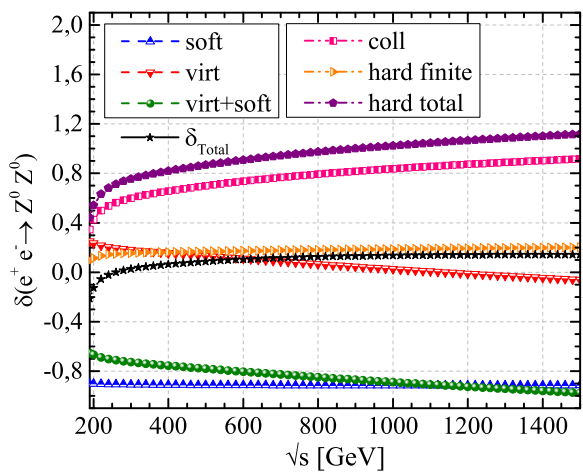}
     \end{center}
     \vspace{-5mm}
\caption{The effect of different individual contributions on cross sections as a function of $\sqrt{s}$.}
\label{fig:delta_corr_type}
\end{figure}
These results show that the pure QED and weak corrections are the same order of magnitude, so that both are equally important. However, the QED correction make the main contribution to the total EW correction. The relative radiative correction due to full EW one-loop contributions in the vicinity $\sqrt{s}$ close to the threshold of production ZZ becomes rather large. This effect is due to the Coulomb singularity in Feynman diagrams, which includes the instantaneous virtual photon exchange in the loop that has a small spatial momentum.  Overall, the full EW one-loop corrections enhance Born cross section. While energy-dependent structure of both the QED and weak corrections is clearly visible, it almost disappears in the total corrections at higher energies. For proposed colliding energies of the future collider projects: CEPC (at $\sqrt{s}=240$ GeV), FCCee (at $\sqrt{s}=350$ GeV), CLIC (at $\sqrt{s}=380$ GeV), and ILC (at $\sqrt{s}=500$ GeV), the unpolarized born-level cross sections reach 1051.44 fb (with $\delta_{\text{Total}}=-2.20\%$), 588.23 fb (with $\delta_{\text{Total}}=+4.73\%$), 513.82 fb (with $\delta_{\text{Total}}=+5.70\%$) and  315.46 fb (with $\delta_{\text{Total}}=+8.66\%$), respectively.
The production rate of $e^- e^+ \to ZZ$ is larger by around one order of magnitude than from the $\gamma\gamma$-collision mode (see Refs.\cite{Gounaris00_3,Bardin17}).

In Fig.~\ref{fig:delta_corr_type}, we plot the virtual, soft and hard photonic corrections as a function of $\sqrt{s}$. The virtual corrections decrease from 0.24 to -0.06, while the soft bremsstrahlung corrections remains nearly constant with a value of -0.91 with the increment of $\sqrt{s}$. However, hard bremsstrahlung corrections, i.e., collinear and non-collinear parts, start from their minimum values and then increase rapidly with the increment of $\sqrt{s}$. Consequently, the virt+soft and hard bremsstrahlung corrections are partially canceled, as they are combined into the full EW corrections. 
\begin{figure}[htb]
    \begin{center}
\includegraphics[scale=0.43]{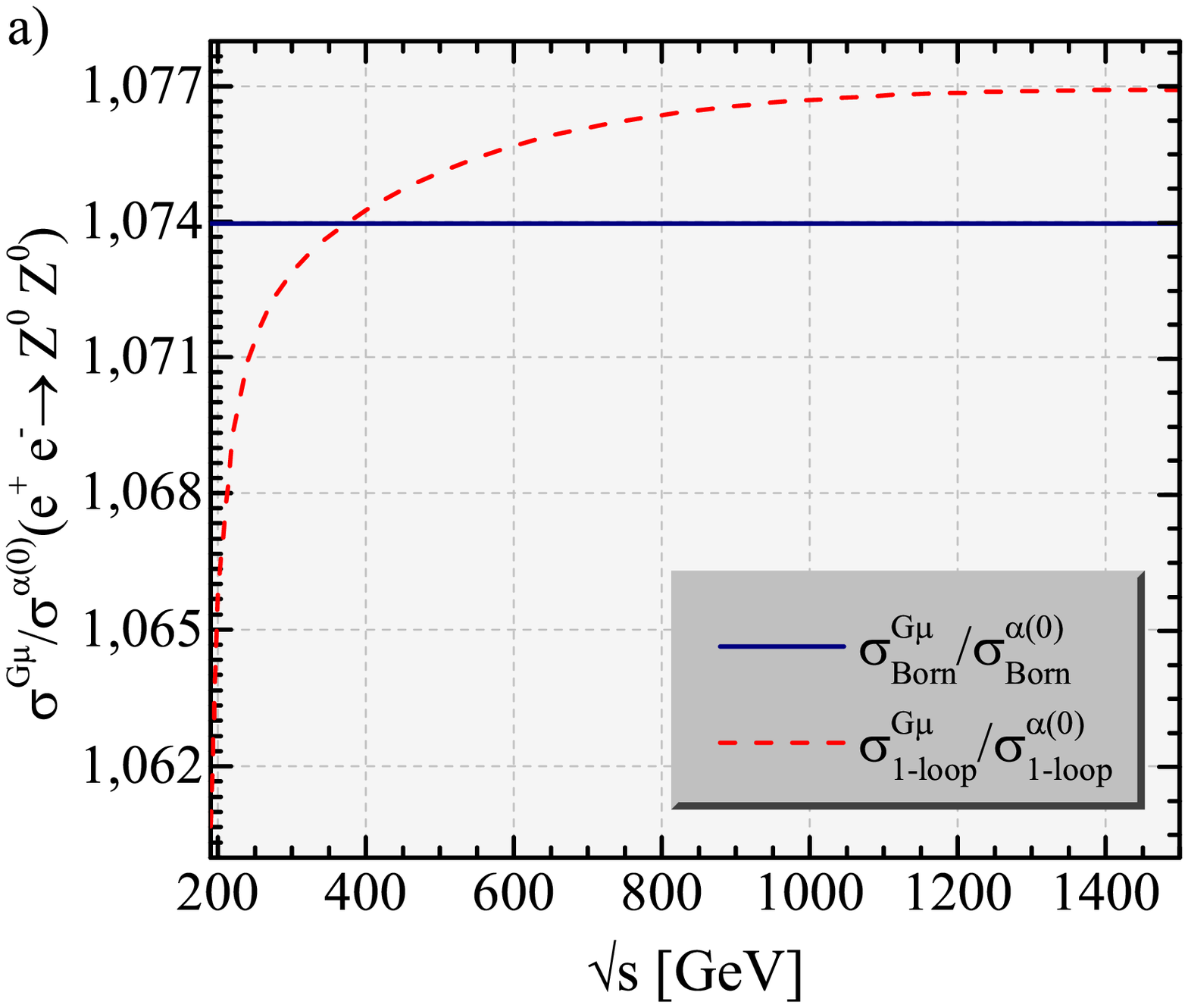}
\includegraphics[scale=0.43]{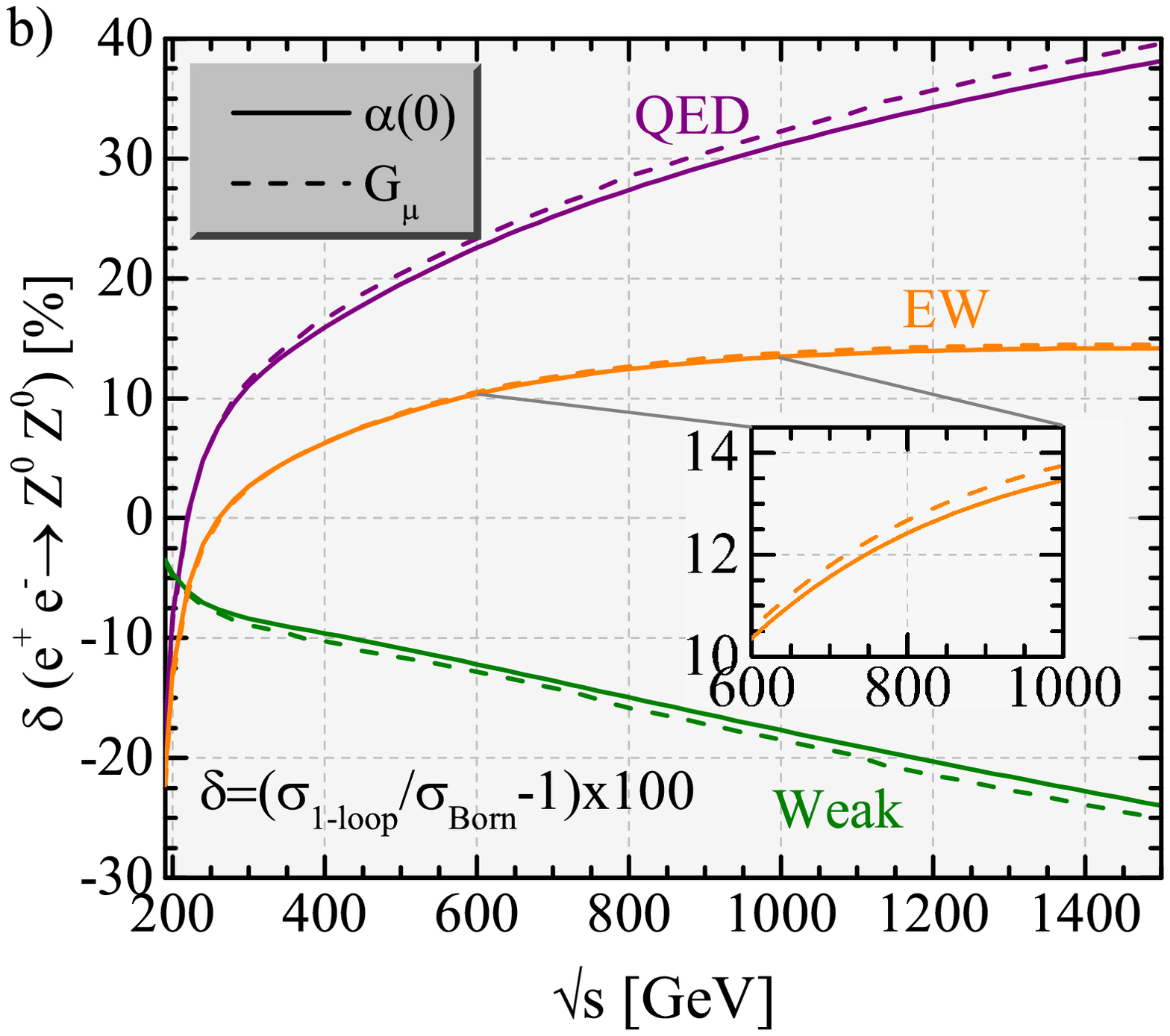}
     \end{center}
     \vspace{-5mm}
\caption{The ratios of cross sections and  relative corrections in two different schemes, $\alpha(0)$ and $G_\mu$ schemes, as a function of $\sqrt{s}$.}
\label{fig:schemes}
\end{figure}
\begin{figure*}[hbt]
    \begin{center}
\includegraphics[scale=0.43]{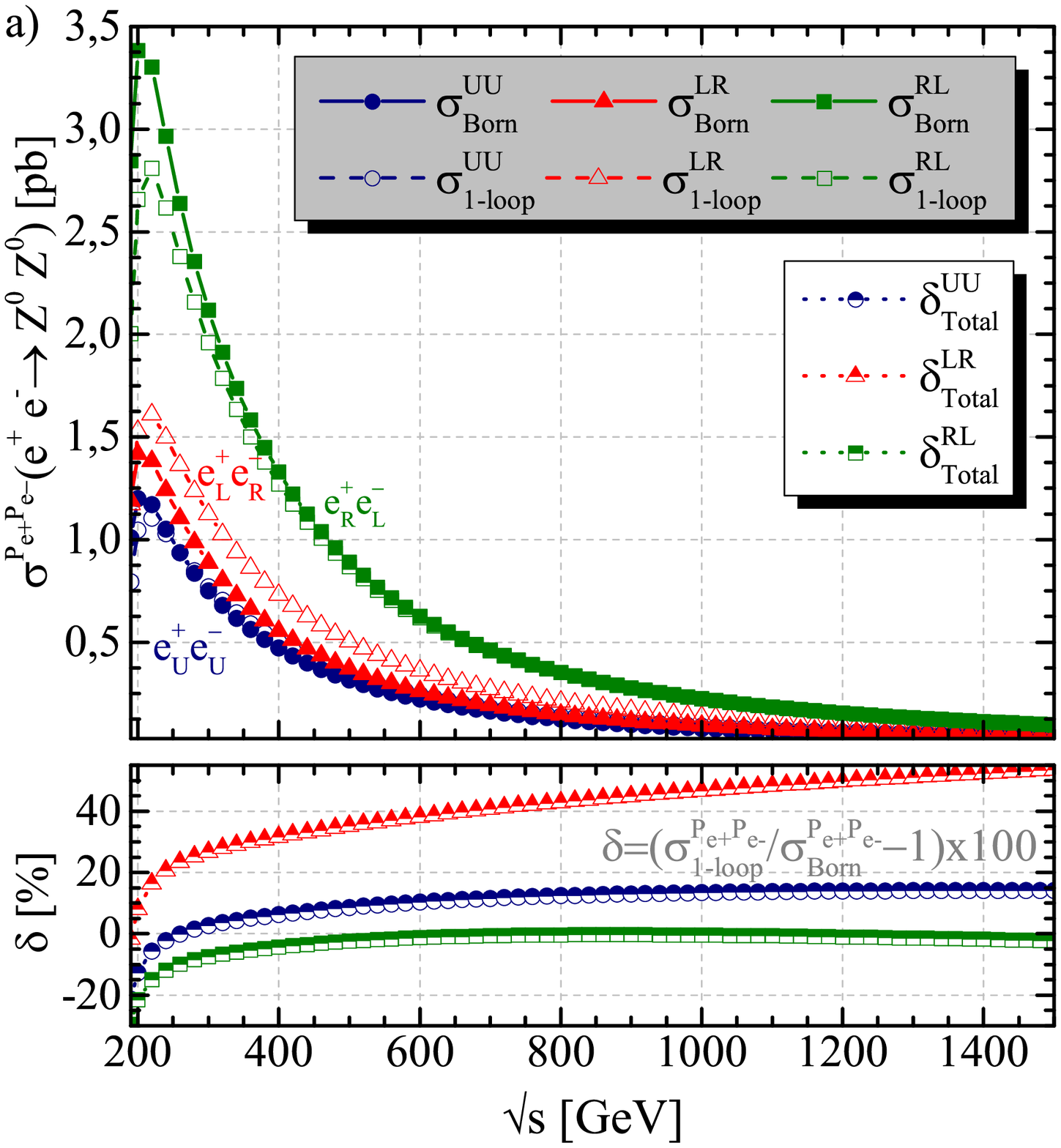}
\includegraphics[scale=0.43]{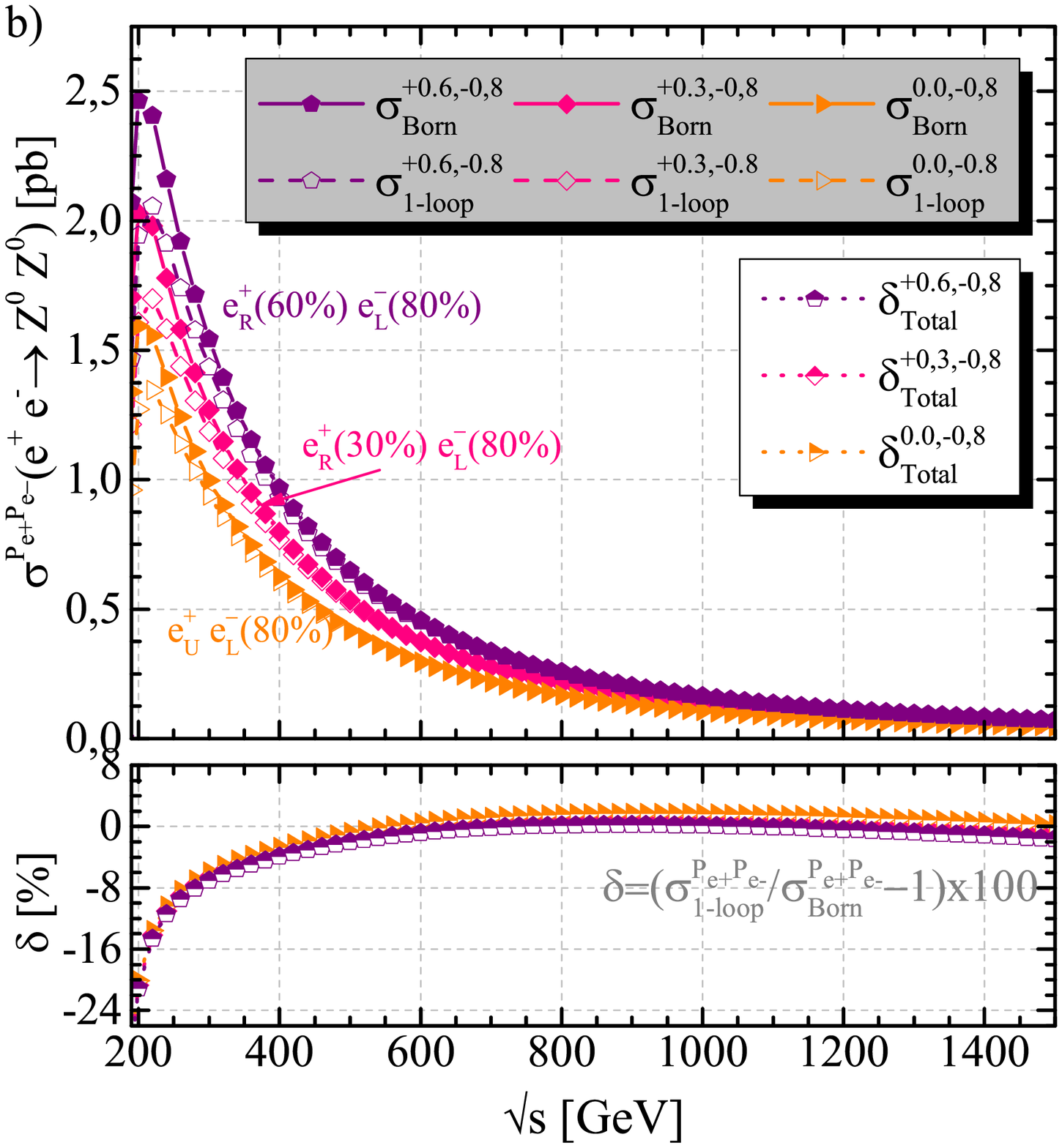}
     \end{center}
     \vspace{-5mm}
\caption{Polarized Born and one-loop cross sections as a function of $\sqrt{s}$.}
\label{fig:pol}
\end{figure*}
The soft and hard bremsstrahlung corrections make the negative and positive dominant contributions to the full EW corrections, respectively.

For $\alpha(0)$ and $G_\mu$ schemes, we present the ratios of cross sections in Fig.~\ref{fig:schemes}(a) and the relative corrections in Fig.~\ref{fig:schemes}(b) as a function of $\sqrt{s}$, respectively. In Table~\ref{table:schemes}, we also provide our numerical results obtained in the $\alpha(0)$ and $G_\mu$ schemes for $\sqrt{s}=250$, $500$, and $1000\gev$. The Born cross section increase by about $7.39\%$ in $G_\mu$ scheme as compared to the $\alpha(0)$ scheme.  When $\sqrt{s}$ goes up from 200 GeV to 1.5 TeV, the one-loop cross section in $G_\mu$ scheme is from $6.58\%$ to $7.69\%$ larger than ones in the $\alpha(0)$ scheme. The QED relative corrections in $G_\mu$ scheme increase by up to about $4\%$ , while the weak relative corrections decrease by up to about $5\%$, as compared to the $\alpha(0)$ scheme. However, the difference between total relative corrections in considered schemes is one percent so small that it can be considered as a theoretical uncertainty.
\begin{table}[t]
\caption{Born and one-loop cross sections, and relative corrections of $e^+ e^- \to Z^0 Z^0$ in the $\alpha(0)$ scheme and $G_\mu$ scheme for various values of $\sqrt{s}$.}\label{table:schemes}
\centering
\begin{ruledtabular}
\begin{tabular}{crrr}
$\sqrt{s}$ & $250\gev$ & $500\gev$ & $1000\gev$\\
\hline
&\multicolumn{3}{c}{$\sigma(e^+ e^- \to Z^0 Z^0)$ [fb]}\\
\cellcolor[gray]{0.7}$\sigma_{\text{Born}}^{\alpha(0)}$   &991.50  &315.46 &79.59 \\
\cellcolor[gray]{0.9}$\sigma_{\text{Born}}^{G_\mu}$       &1064.83 &338.88  &85.48 \\
\cellcolor[gray]{0.7}$\sigma_{\text{1-loop}}^{\alpha(0)}$ &981.59  &342.77 &90.31 \\
\cellcolor[gray]{0.9}$\sigma_{\text{1-loop}}^{G_\mu}$     &1051.67 &368.51 &97.24 \\
&\multicolumn{3}{c}{$\delta(e^+ e^- \to Z^0 Z^0)$ [$\%$]}\\
\cellcolor[gray]{0.7}$\delta_{\text{QED}}^{\alpha(0)}$   &+6.35  &+19.50 &+31.15 \\
\cellcolor[gray]{0.9}$\delta_{\text{QED}}^{G_\mu}$       &+6.39   &+20.38  &+32.27 \\
\cellcolor[gray]{0.7}$\delta_{\text{Weak}}^{\alpha(0)}$  &$-$7.35  &$-$10.85 &$-$17.69 \\
\cellcolor[gray]{0.9}$\delta_{\text{Weak}}^{G_\mu}$      &$-$7.63   &$-$11.61 &$-$18.52 \\
\cellcolor[gray]{0.7}$\delta_{\text{Total}}^{\alpha(0)}$ &$-$1.00  &+8.65 &+13.46 \\
\cellcolor[gray]{0.9}$\delta_{\text{Total}}^{G_\mu}$     &$-$1.24   &+8.77  &+13.75 \\
\end{tabular}
\end{ruledtabular}
\end{table}

Now we examine the initial beam polarization dependence of the Born and the one-loop cross sections ($\sigma^{P_{e^{+}}P_{e^{-}}}_\text{Born}$ and $\sigma^{P_{e^+}P_{e^-}}_\text{1-loop}$ ) on $\sqrt{s}$. Also, we present the total relative corrections in order to see effect of polarization configurations on the EW corrections. Figure~\ref{fig:pol}(a) shows unpolarized and completely polarized initial beams cases. Here we use the following notation: $\sigma^\text{RL}$ denotes the total cross section with the $100\%$ right-handed polarized positron ($P_{e^+}=+1$), and the $100\%$ left-handed polarized electron ($P_{e^-}=-1$) beams, $({e^+_R},{e^-_L})$. Others can be defined analogously. We note that $\sigma^\text{LL}$ and $\sigma^\text{RR}$ are very small ($\mathcal{O}(10^{-10})$  pb) as expected, and are thus not included here. All curves for polarized and unpolarized cases increase quickly with the opening of the phase space, reach a maximum value and then fall off rapidly with the increment of $\sqrt{s}$. Their maximum values are reached around at $\sqrt{s}\sim 210$ GeV. The one-loop polarized cross sections $\sigma^{\text{LR}}_\text{1-loop}$ and $\sigma^{\text{RL}}_\text{1-loop}$ can be enhanced by about a factor of 1.5 and 2.5, respectively, compared with the unpolarized case. For instance, at $\sqrt{s}=$ 250 GeV, $\sigma^\text{RL}_\text{Born}$ reaches a value of 2.80 pb, yielding a total relative correction of about $-10.71\%$. At $\sqrt{s}=$ 250 GeV, $\sigma^\text{LR}_\text{Born}$ reaches a value of 1.17 pb, yielding a total relative correction of $+22.20\%$. While $\sigma^\text{RL}_\text{Born}$ is around 2 times larger than $\sigma^\text{RL}_\text{Born}$, the relative corrections of the former are smaller than that of the latter. Namely, $({e^+_L}{e^-_R})$ polarization case has larger EW radiative corrections than other cases.

\begin{figure*}[!t]
    \begin{center}
\includegraphics[scale=0.43]{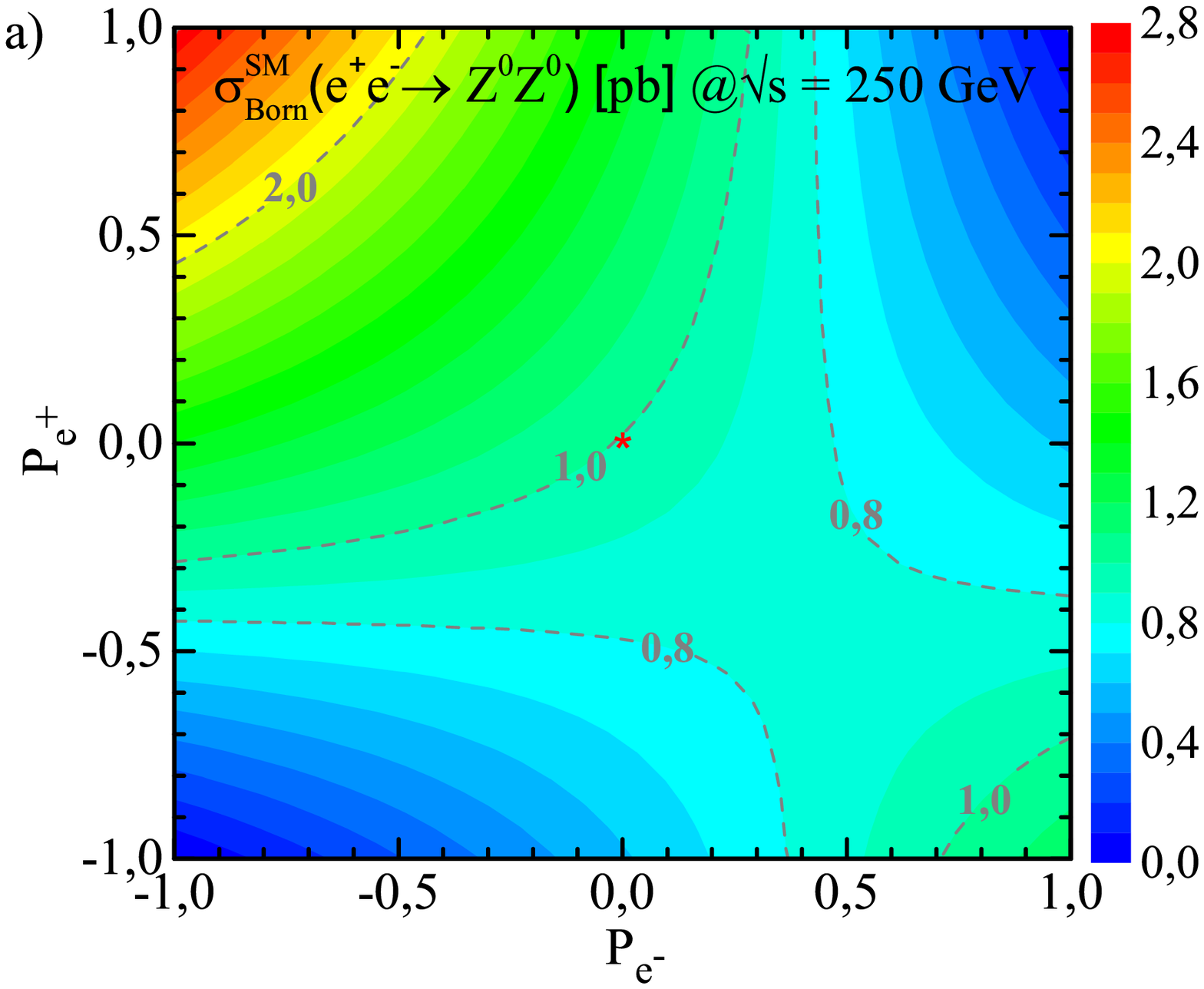}
\includegraphics[scale=0.43]{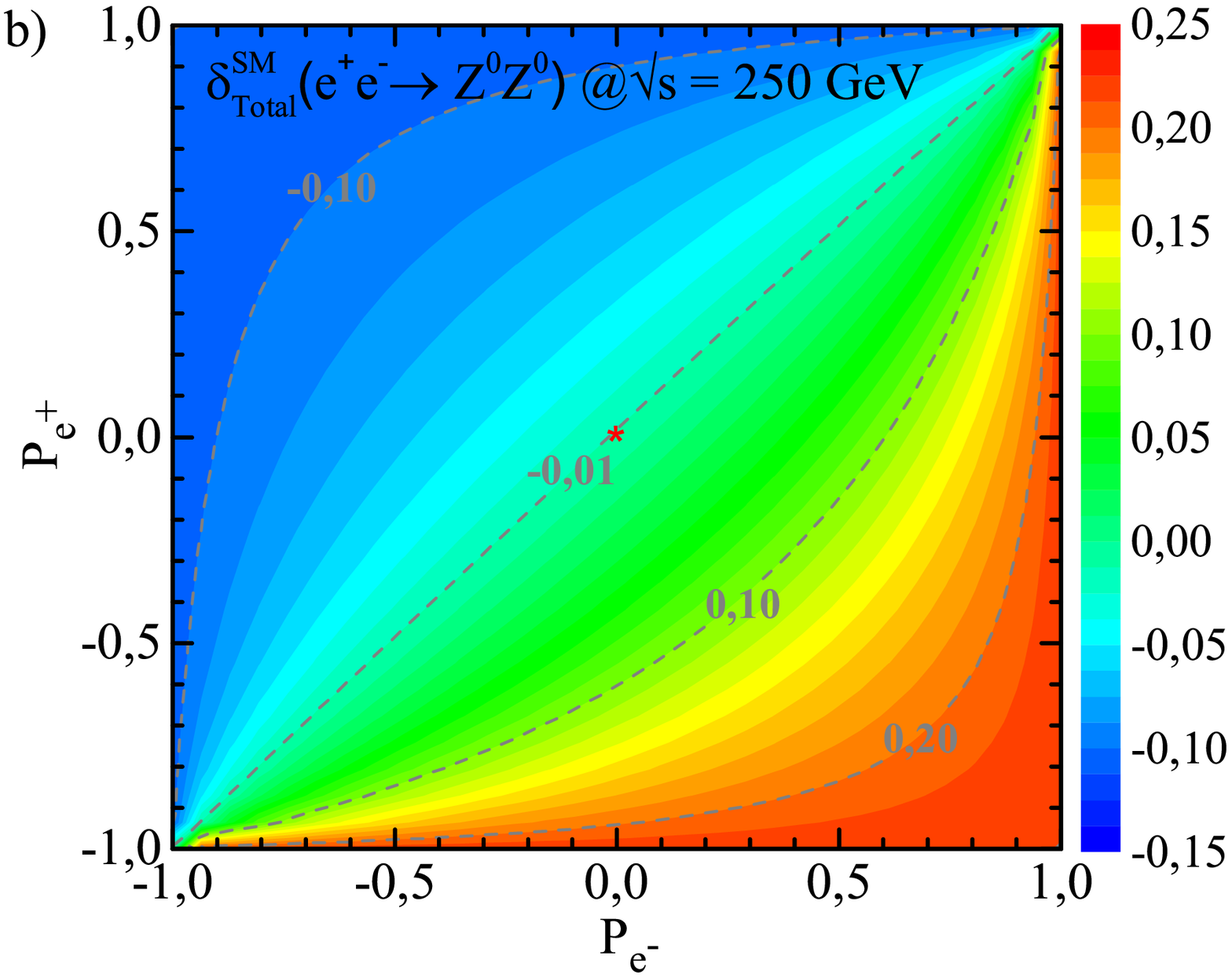}
     \end{center}
     \vspace{-5mm}
\caption{(color online). Beam polarization effects on the Born cross section and on the total relative correction for $\sqrt{s}=250\gev$. The color heat maps correspond to the values of born cross section (left plot) and the total EW relative correction (right plot). The asterisk denotes the unpolarized point $(P_{e^+}, P_{e^-})=(0,0)$. }
\label{fig:pol2D250}
\end{figure*}
\begin{figure*}[ht]
    \begin{center}
\includegraphics[scale=0.43]{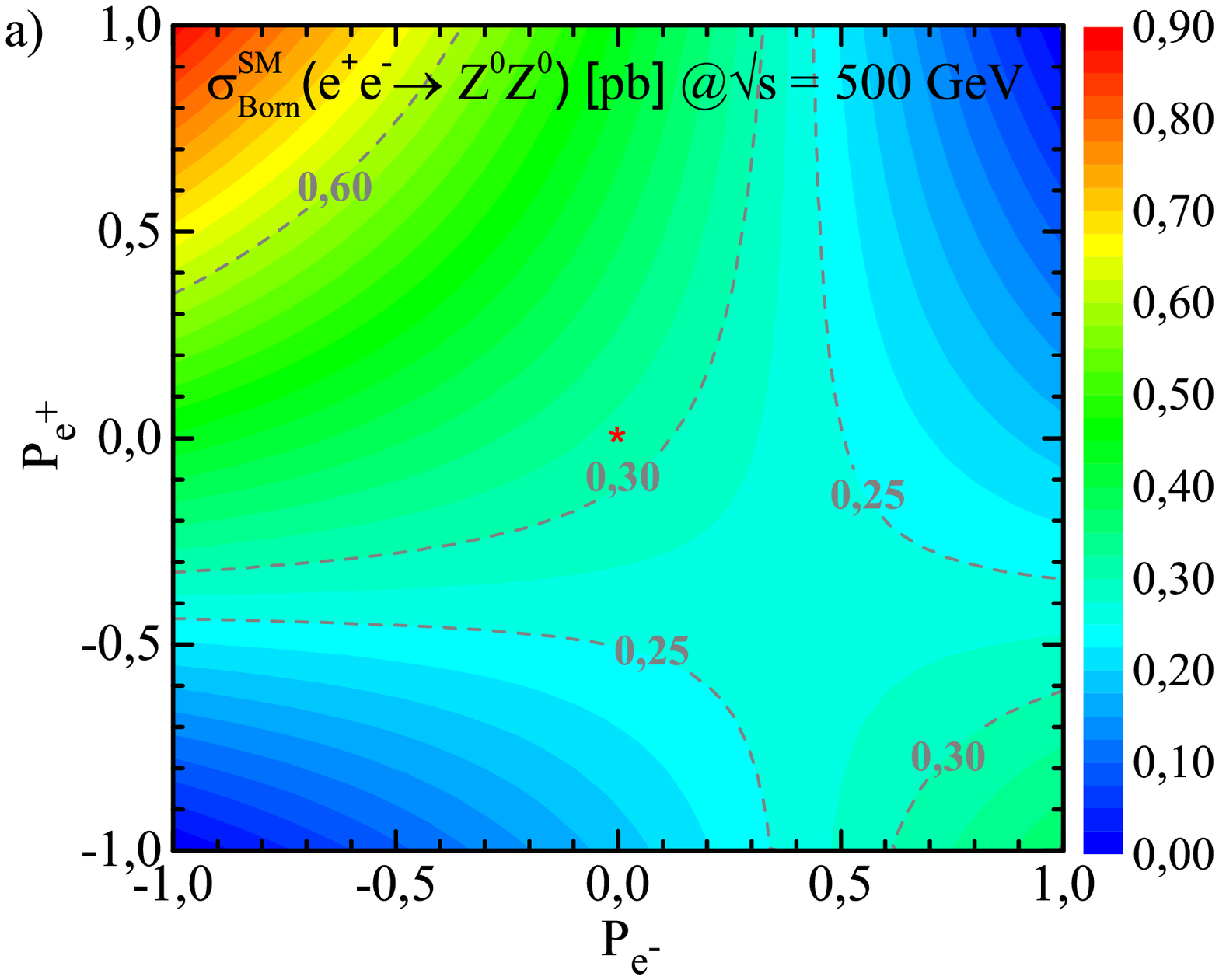}
\includegraphics[scale=0.43]{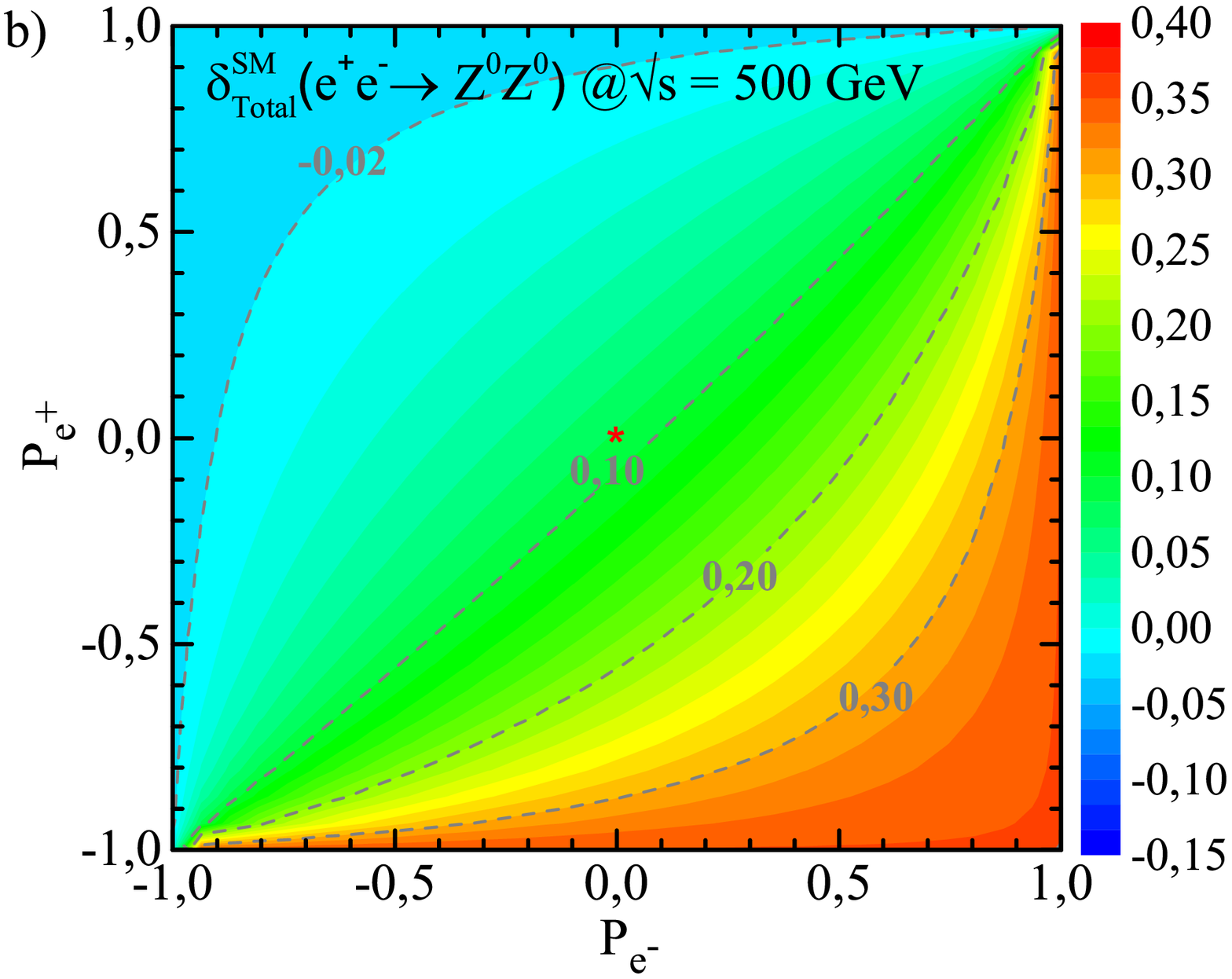}
     \end{center}
     \vspace{-5mm}
\caption{(color online). Same as in Fig.~\ref{fig:pol2D250} but for $\sqrt{s}=500\gev$.}
\label{fig:pol2D500}
\end{figure*}
Figure~\ref{fig:pol}(b) shows the results for various polarization degrees of the initial beams: $(P_{e^+}, P_{e^-})=(0.0,-0.8), (+0.3, -0.8), (+0.6,-0.8)$, proposed by the future colliders. The cross sections at both Born and one-loop level are sorted according to various polarization degrees of initial beams as follows: $\sigma^{\text{RL}}>\sigma^{\text{+0.6,-0.8}}>\sigma^{\text{+0.3,-0.8}}>\sigma^{\text{0.0,-0.8}}>\sigma^{\text{LR}}>\sigma^{\text{UU}}$. There appears a similar energy dependence behavior in polarized cases as in the unpolarized case. The one-loop cross section with the polarization degrees of $(P_{e^+}, P_{e^-})=(+0.6,-0.8)$, i.e., $\sigma^{(+0.6,-0.8)}_{\text{1-loop}}$, has a maximum of 2.05 pb, providing a total relative correction of $-14.59\%$. When $\sqrt{s}$ goes from 240 GeV to 1.0 TeV, the relative corrections vary from about $-11\%$ to $+1.69\%, +0.75\%$ and $+0.14\%$ for $(P_{e^+}, P_{e^-})=(0.0,-0.8), (+0.3, -0.8), (+0.6,-0.8)$, respectively.
We obtain the following results for proposed polarization cases by the future collider projects: $\sigma^{\text{+0.3,-0.8}}_{\text{1-loop}}=$ 1.51 pb with $\delta^{\text{+0.3,-0.8}}_{\text{Total}}=-9.90\%$ at $\sqrt{s}=250$ GeV (ILC) and  $\sigma^{\text{0,-0.8}}_{\text{1-loop}}=$ 0.66 pb with $\delta^{\text{0,-0.8}}_{\text{Total}}=-3.94\%$ at  $\sqrt{s}=380$ GeV (CLIC), respectively.

In Figs.~\ref{fig:pol2D250} and \ref{fig:pol2D500}, the Born cross section and the total relative correction are also presented in the plane of polarization degrees of the incoming beams $(P_{e^+}, P_{e^-})$ for $\sqrt{s}=250\gev$ and $\sqrt{s}=500\gev$, respectively. The $P_{e^+}$ and $P_{e^-}$ range from $-1$ to $+1$. We also show some values by the contour lines. The beam polarization dependence of the relative correction is calculated from $\delta^{P_{ e^+}P_{ e^-}}_{\text{Total}}=\bigl(\sigma^{P_{ e^+}P_{ e^-}}_\text{1-loop}/\sigma^{P_{ e^+}P_{ e^-}}_\text{Born}-1\bigr)$ by using Eq.~\eqref{eq:polsigma}.
The Born cross section reaches its larger values at the left top corner ( $0 < P_{e^+} \leq +1$ and $ -1 \leq P_{e^-} < 0$), whereas it has smaller values at the right top and left bottom corners. As excepted, it has a maximum value at point $(P_{e^+}, P_{e^-})=(+1,-1)$, namely, $100\%$ left-handed polarized electron and $100\%$ right-handed polarized positron. The total relative correction reaches positive and larger values at the right bottom corner ($-1 \leq P_{e^+} < 0$ and $ 0 < P_{e^-} \leq +1$). As the values of polarization degrees approach to $P_{e^-} \rightarrow +1$ and $P_{e^+} \rightarrow -1$, the relative correction increases. Particularly, significant positive corrections are observed in the region below the 0.10 contour line. Together with the larger cross sections, smaller systematic errors can be expected for the cross-section measurement with the polarized beams than in the unpolarized case.

\begin{figure}[!t]
    \begin{center}
\includegraphics[scale=0.43]{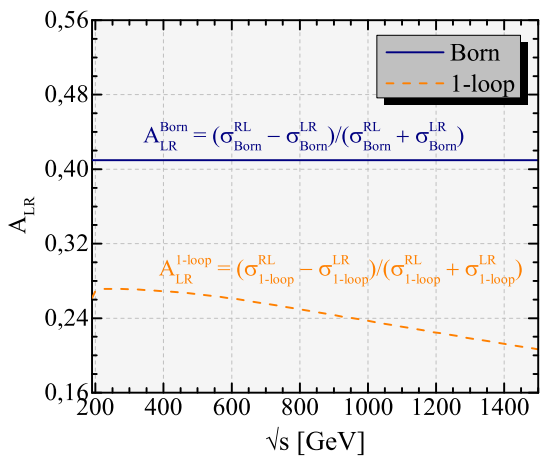}
     \end{center}
     \vspace{-5mm}
\caption{The left-right asymmetry as a function of $\sqrt{s}$.}
\label{fig:asym}
\end{figure}

\begin{table*}[th]
\caption{Born$\&$one-loop cross sections, relative correction and left-right asymmetry for various polarization configurations at $\sqrt{s}=250$. Here, we define the scaling factor as $\text{R}^{P_{ e^+}P_{ e^-}}_X=\sigma^{P_{ e^+}P_{ e^-}}_\text{X}/\sigma^{0,0}_\text{X}$, where $X$ stands for Born or one-loop level.}\label{tab:pol}
\centering
\begin{ruledtabular}
\begin{tabular}{cccccrrrcc}
$(P_{e^+}, P_{e^-})$ &$P_{eff}$ &$L_{eff}/L$ & $A_{LR}^{\text{obs}}(\text{Born})$& $A_{LR}^{\text{obs}}(\text{1-loop})$&$\sigma_{\text{Born}}$  [fb]& $\sigma_{\text{1-loop}}$  [fb]& $\delta_{\text{Total}} [\%]$& $\text{R}^{P_{ e^+}P_{ e^-}}_\text{Born}$& $\text{R}^{P_{ e^+}P_{ e^-}}_\text{1-loop}$\\
\hline
$(~0,~0)$ &$0$  &0.50&0 &0&991.50   &981.59   &$-0.99\%$  &1.00  &1.00 \\
$(+1.0,-1.0)$   &$-1.00$ &1.00&0.4098 &0.2715&2795.61  &2496.35  &$-10.71\%$ &2.82  &2.54 \\
$(-1.0,+1.0)$   &$+1.00$ &1.00&0.4098 &0.2715&1170.39  &1430.25  &$+22.20\%$ &1.18  &1.46 \\
$(~~0.0,-1.0)$  &$-1.00$ &0.50&0.4098 &0.2715&1397.81  &1248.17 &$-10.70\%$  &1.41  &1.27 \\
$(~~0.0,-0.8)$  &$-0.80$ &0.50&0.3278 &0.2172&1316.54  &1194.87  &$-9.24\%$  &1.33  &1.22 \\
$(~~0.0,+0.8)$  &$+0.80$ &0.50&0.3278 &0.2172&666.46   &768.43  &$+15.30\%$  &0.67  &0.78 \\
$(+0.3,-0.8)$   &$-0.89$ &0.62&0.3635 &0.2409&1676.40  &1510.42  &$-9.90\%$  &1.69  &1.54 \\
$(-0.3,+0.8)$   &$+0.89$ &0.62&0.3635 &0.2409&782.52   &924.07  &$+18.09\%$  &0.79  &0.94 \\
$(+0.6,-0.8)$   &$-0.95$ &0.74&0.3876 &0.2568&2036.25  &1825.98  &$-10.33\%$ &2.05  &1.86 \\
$(-0.6,+0.8)$   &$+0.95$ &0.74&0.3876 &0.2568&898.59   &1079.71  &$+20.16\%$ &0.91  &1.10 \\
\end{tabular}
\end{ruledtabular}
\end{table*}
A convenient observable is the left–right asymmetry when applying polarized beams. This asymmetry is especially important for high precision measurements. Therefore, we present the left-right asymmetry of the cross sections, which is defined in Eq.~\eqref{eq:ALR}, as a function of center-of-mass energy in Fig.~\ref{fig:asym}. For Born-level, the left-right asymmetry remains a stable of 0.40979 with the increment of $\sqrt{s}$. This value is exactly the same as the result calculated theoretically in Eq.~\eqref{eq:ALRBORN}.  On the other hand, at one-loop level, the left-right asymmetry decrease from 0.2616 to 0.2065, as the center-of-mass energy goes up from 190 to 1500 GeV. Consequently, the radiative corrections can conveniently be described by the one-loop induced deviation from its value at Born-level.

\begin{figure}[t]
    \begin{center}
\includegraphics[scale=0.43]{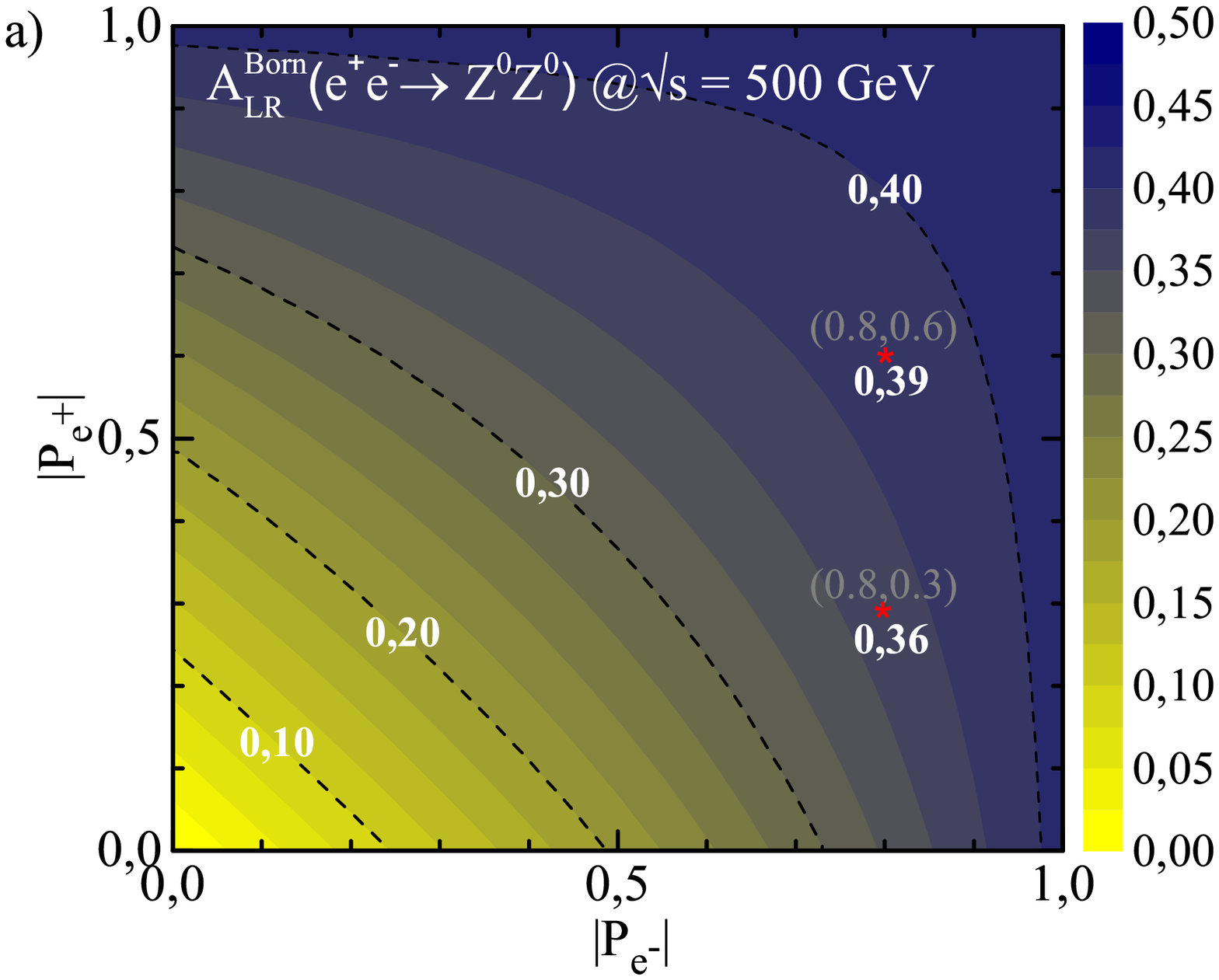}
\includegraphics[scale=0.43]{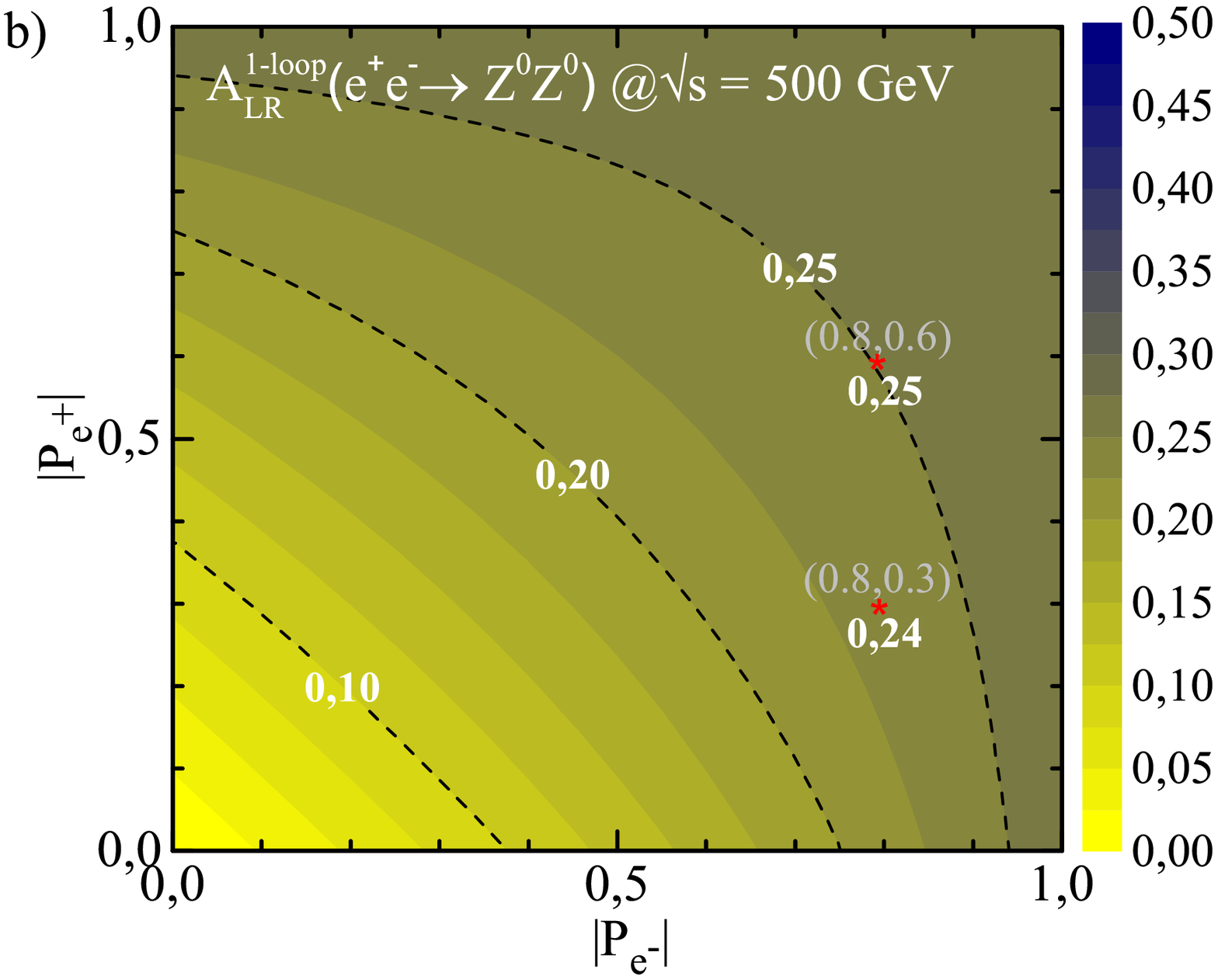}
     \end{center}
     \vspace{-5mm}
\caption{(color online). The left-right asymmetry in the plane of polarization degrees of the incoming beams $(P_{e^+}, P_{e^-})$ for $\sqrt{s}=500$.}
\label{fig:asym2D}
\end{figure}

In Figs.~\ref{fig:asym2D}(a)-(b), we plot the left-right asymmetry of the cross sections $A_{LR}^{\text{obs}}$, which is defined in Eq.~\eqref{eq:ALRobs}, as a function of polarization degrees of $(P_{e^+}, P_{e^-})$. Some values are also shown with contour lines. We note that the results are symmetric with respect to the exchange of $P_{e^+}\leftrightarrow P_{e^-}$, since they only appear by their absolute values in the Eq.~\eqref{eq:sigmapm}. It is obvious that the left-right asymmetry $A_{LR}^{\text{obs}}$ increases as both $|P_{e^+}|$ and $|P_{e^-}|$ run from 0 to 1. The maximum values reaches at the point $(|P_{e^+}|,|P_{e^-}|)=(1,1)$. These values are 0.40979 and 0.2657 for Born and one-loop level, respectively. In these figures, we also mark some special points by asterisk.

Table~\ref{tab:pol} presents the numerical values of the polarized cross sections, the relative corrections and the left-right asymmetry $A_{LR}^{\text{obs}}$ for various polarization degrees of the initial beams.
\begin{figure*}[!thb]
    \begin{center}
\includegraphics[scale=0.40]{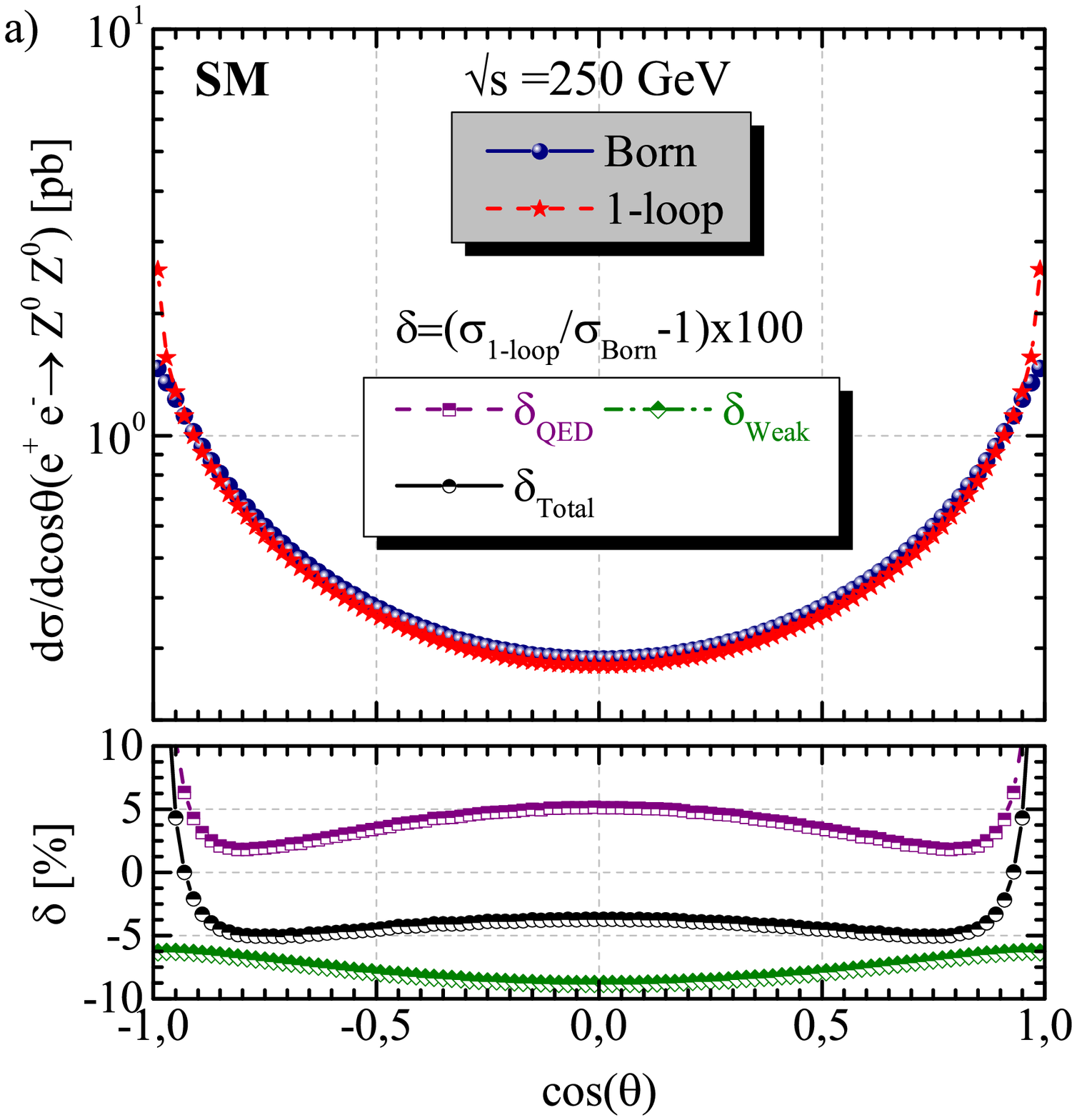}
\includegraphics[scale=0.40]{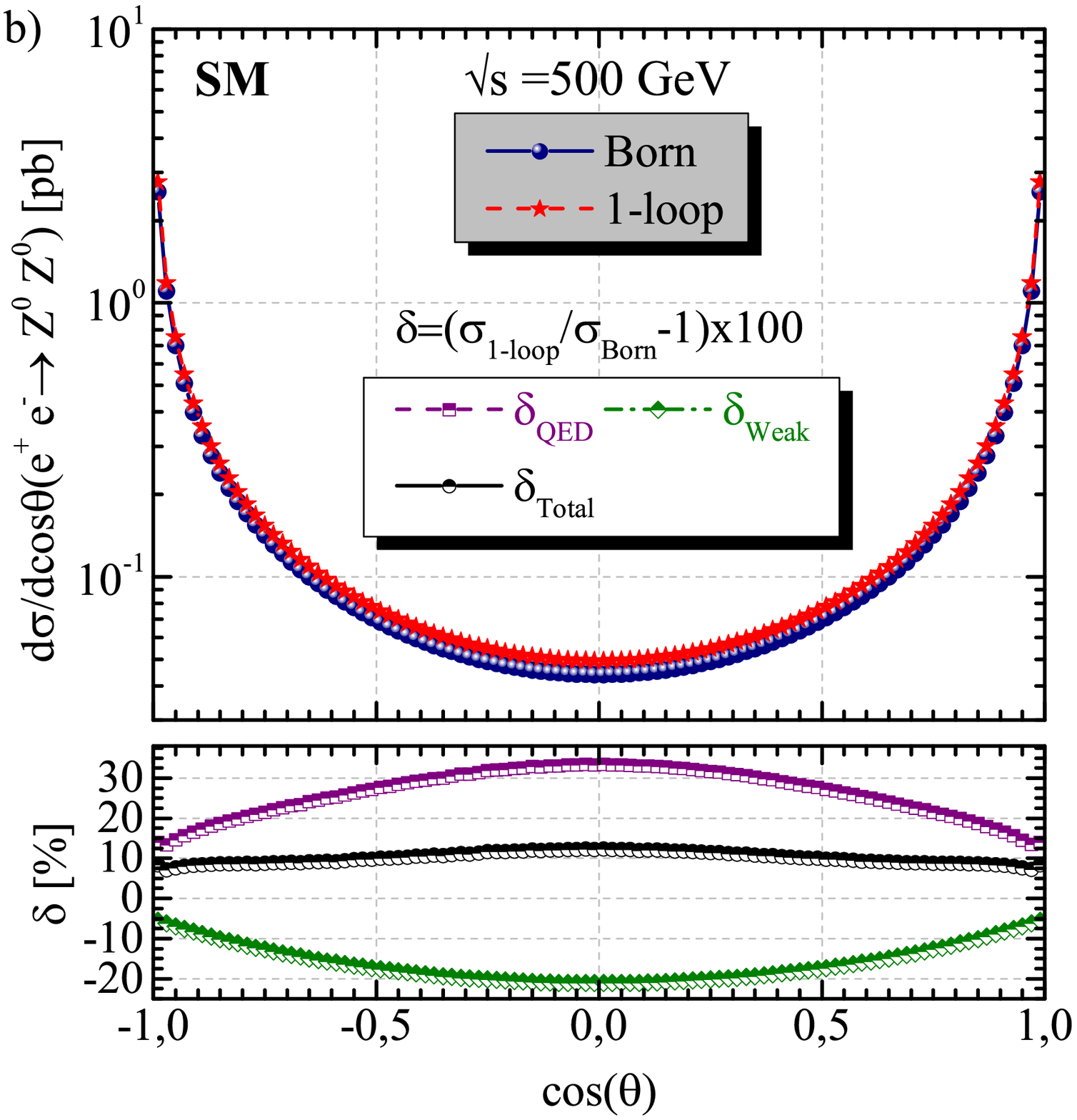}
     \end{center}
     \vspace{-5mm}
\caption{Born and one-loop angular distributions for (a) $\sqrt{s}=250\gev$ and (b) $\sqrt{s}=500\gev$.}
\label{fig:difangular}
\end{figure*}
The values of effective polarization and effective luminosity, which are calculated by Eqs.~\eqref{eq:Peff} and~\eqref{eq:Leff}, respectively, are also given. It is clear that the production rate can be enhanced if $P_{e^-}$ and $P_{e^+}$ have negative and positive signs, respectively. In particular, the longitudinally $(P_{e^+}, P_{e^-})=(+1,-1)$ polarizations of the initial beams significantly improve the cross-section.

Figures~\ref{fig:difangular}(a) and (b) show the differential cross sections at Born and one-loop level as a function of the angle between the coming electron and the outing Z boson for $\sqrt{s}=250$ and $500\gev$, respectively. The angular dependence of the QED, weak and total EW relative corrections is also shown in the same figures.
For Born and one-loop cross sections, the angular distribution peaks significantly in the (symmetrically) forward and backward directions. The relative corrections modify somewhat the Born angular distribution because their influence is larger in the central region. The total corrections reach their maximum values, when $\cos{\theta}$ goes to the extreme points -1 or +1 for $\sqrt{s}=250$ and zero point for $\sqrt{s}=500$, respectively. Namely, the Z bosons are dominantly produced in the forward and backward directions for $\sqrt{s}=250$, whereas in the central region for $\sqrt{s}=500$. Therefore, it will be more likely to observe them in these collision regions. For all values of $\cos{\theta}$, the QED corrections make a positive contributions, while the weak corrections supply a negative contributions. The same behavior is observed as in the total cross sections. The QED and weak contributions are partially offset by each other into EW contributions. While both corrections show a significant dependence on $\cos{\theta}$ separately, this decreases for full EW corrections. When $\cos{\theta}$ runs from 0 to +0.95 or -0.95, at $\sqrt{s}=250\gev$, the total relative correction $\delta_{\text{Total}}$ increases from $-3.71\%$ to $4.30\%$, while decreases $12.2\%$ to $7.7\%$ at $\sqrt{s}=500\gev$.

\section{Summary and Conclusions}\label{sec:conc}
For further sensitivity testing of the SM, as well as to look for clues on the BSM, high precision calculations should be performed. At least, a full set of one-loop corrections in the production channels must be included to ensure adequate accuracy. However, the extent to which higher-order computations beyond one-loop order will be needed depends largely on the expected experimental accuracy.

In this work, by considering a full set of one-loop EW corrections, we have investigated the Z-boson pair production at electron-positron collisions. The UV divergences have been organized by dimensional regularization on the OS scheme. Furthermore, the IR divergences have been removed by the inclusion of the soft and hard bremsstrahlung corrections. Also, the collinear divergences have been corrected by the phase space slicing method. We have verified the stability of our results against the angular and soft cutoff parameters $\Delta_c$, $\Delta_s$, as well as the IR regulator $m_\gamma$. We have also compared the results from \textsc{Whizard} and \textsc{CalcHEP} for the Born and hard photon bremsstrahlung cross sections with the results of the packages used in this work, and obtained a very good agreement (up to six digits).

We have carried out the numerical evaluation for the $\alpha(0)$ and $G_\mu$ schemes. The differences between these schemes for the QED and weak relative corrections are about $4\%$ and $5\%$, respectively, whereas this is $1\%$  for the total relative correction. This can be
considered as a theoretical uncertainty.

Our results show that  the Born cross section is commonly increased by the one-loop EW radiative corrections and the total relative correction is typically up to about ten percent. The pure QED and weak corrections are the same order of magnitude, so that both are important for precision calculation. The QED and weak contributions are partially offset by each other into EW contributions. However, the QED correction makes the main contribution to the total EW correction.

Moreover, we have investigated the spin polarization effects of the initial beams on the total cross sections. We have observed that the radiative corrections have a large polarization dependence. As a result, an improvement has been observed by a factor of 2.5 with the $100\%$ right-handed polarized positron and the $100\%$ left-handed polarized electron beams, compared with the unpolarized case. The left-right asymmetry and angular distributions have also presented. The Born and one-loop angular distributions are symmetric and strongly peaked in the forward and  backward directions. For all angles, the QED corrections make a positive contributions, while the weak corrections supply a negative contributions. The relative corrections slightly change the Born angular distribution as their effect is usually larger in the central region.

In summary, an analysis of the one-loop EW radiative corrections to Z boson-pair production in electron-positron collisions have been carried out in the framework of the SM.  The initial beam effects on the cross sections have been discussed in detail. It has been clearly shown that one-loop EW radiative corrections significantly alter the lowest-order results and should therefore be fully accounted for a realistic description of experiments at future collider energies. Our results provide precise predictions for Z-boson pair production, which can be tested as experiments achieve higher accuracy.

\begin{acknowledgments}
The work of M. Demirci was supported by the Scientific and Technological Research Council of Turkey (TUBITAK) in the framework of 2219-International Postdoctoral Research Fellowship Program. This work was also supported in part by the National Science Foundation Grant No. PHY-2108339.
\end{acknowledgments}


\begin{thebibliography}{99}
\bibitem{Glashow61} S. L. Glashow, 
 \href{https://doi.org/10.1016/0029-5582(61)90469-2}{\textit{Nucl. Phys.}~{\bf 22}, 579-588 (1961)}.

\bibitem{Weinberg67} S. Weinberg, 
 \href{https://doi.org/10.1103/PhysRevLett.19.1264}{\textit{Phys. Rev. Lett.}~{\bf 19}, 1264-1266 (1967)}.

\bibitem{Salam68} A. Salam, 
 \href{https://doi.org/10.1142/9789812795915 0034}{\textit{Conf. Proc. C}~{\bf 680519}, 367-377 (1968)}.

\bibitem{ATLAS} G. Aad {\it et al.} (ATLAS Collaboration), \href{https://doi.org/10.1016/j.physletb.2012.08.020}{\textit{Phys. Lett. B}~{\bf 716}, 1 (2012)}.
\bibitem{CMS} S. Chatrchyan {\it et al.} (CMS Collaboration), \href{https://doi.org/10.1016/j.physletb.2012.08.021}{\textit{Phys. Lett. B}~{\bf 716}, 30 (2012)}.
\bibitem{ATLASCMS} G. Aad {\it et al.} (ATLAS and CMS Collaborations), \href{http://dx.doi.org/10.1007/JHEP08(2016)045}{\textit{J. High Energ. Phys.}~\textbf{2016}, 45 (2016)}.

\bibitem{ILC1} P.~Bambade {\it et al.}, {\it The International Linear Collider: A Global Project},    Report No. DESY 19-037,
   \href{https://doi.org/10.48550/arXiv.1903.01629}{ arXiv:1903.01629 [hep-ex]}.
\bibitem{ILC2} H.~Baer \textit{et al.}, {\it Physics Chapter of the ILC Detailed Baseline Design Report}, Report No. ILC-REPORT-2013-040,
   \href{https://doi.org/10.48550/arXiv.1306.6352}{arXiv:1306.6352 [hep-ph]}.

\bibitem{ILC3} H. Aihara {\it et al.}, {\it  The International Linear Collider A Global Project},
   \href{https://doi.org/10.48550/arXiv.1901.09829}{arXiv:1901.09829 [hep-ex]}.

\bibitem{CLIC1} E. Accomando \textit{et al.}~(CLIC Physics Working Group), {\it Physics at the CLIC Multi-TeV Linear Collider}, Report No. CERN-2004-005,
   \href{https://doi.org/10.48550/arXiv.hep-ph/0412251}{arXiv:hep-ph/0412251}.

\bibitem{CEPC} The CEPC Study Group,  {\it CEPC Conceptual Design Report: Volume 2 - Physics $\&$ Detector}, Report No. IHEP-CEPC-DR-2018-02, 2018,
   \href{https://arxiv.org/abs/1811.10545}{arXiv:1811.10545 [hep-ex]}.


\bibitem{FCC19} M. Benedikt \textit{et al.},  {\it Future Circular Collider - European Strategy Update Documents: The FCC integrated programme (FCC-int)}, Report No. CERN-ACC-2019-0007, https://cds.cern.ch/record/2653673, Jan, 2019.

\bibitem{FCCee} A. Abada \textit{et al.} 
    \href{https://doi.org/10.1140/epjst/e2019-900045-4}{\textit{Eur. Phys. J. Spec. Top.}~{\bf  228}, 261 (2019)}.


\bibitem{Pankov06} A. A. Pankov, A. V. Tsytrinov, and N. Paver, 
    \href{https://link.aps.org/doi/10.1103/PhysRevD.73.115005}{\textit{Phys. Rev. D}~{\bf 73},115005 (2006)}.

\bibitem{Osland10} P. Osland, A. A. Pankov, and  A. V. Tsytrinov, 
    \href{https://doi.org/10.1140/epjc/s10052-010-1272-z}{\textit{Eur. Phys. J. C}~{\bf 67}, 191–204 (2010)}.

\bibitem{Moortgat08} G. Moortgat-Pick {\it et al.} 
    \href{https://doi.org/10.1016/j.physrep.2007.12.003}{\textit{Phys. Rep.}~{\bf 460}, 131-243 (2008)}.

\bibitem{Brown79} R. W. Brown and K. O. Mikaelian, 
    \href{https://link.aps.org/doi/10.1103/PhysRevD.19.922}{\textit{Phys. Rev. D}~{\bf 19}, 922 (1979)}.

\bibitem{Gaemers79} K. J. F. Gaemers and G. J.  Gounaris, 
    \href{https://doi.org/10.1007/BF01440226}{\textit{Z. Phys. C - Particles and Fields }~{\bf 1}, 259–268 (1979)}.

\bibitem{Denner88} A. Denner and T. Sack, 
    \href{https://doi.org/10.1016/0550-3213(88)90691-8}{\textit{Nucl. Phys. B}~{\bf 306}, 221 (1988)}.


\bibitem{Gounaris03} G. J. Gounaris, J. Layssac, and  F. M. Renard, 
\href{https://link.aps.org/doi/10.1103/PhysRevD.67.013012}{\textit{Phys. Rev. D}~{\bf 67}, 013012 (2003)}.

\bibitem{OPAL} G. Abbiendi {\it et al.} (OPAL Collaboration), \href{https://doi.org/10.1016/S0370-2693(00)00197-0}{\textit{Phys. Lett. B}~{\bf 476}, 256 (2000)}.

\bibitem{Exp03} LEP, ALEPH, DELPHI, L3, OPAL, LEP Electroweak Working Group, SLD Electroweak Group, SLD Heavy Flavor Group Collaboration, {\it A Combination of preliminary electroweak measurements and constraints on the standard model},
       \href{https://doi.org/10.48550/arXiv.hep-ex/0312023}{arXiv:hep-ex/0312023 [hep-ex]}.

\bibitem{L3} P. Achard {\it et al.} (L3 Collaboration), \href{https://doi.org/10.1016/j.physletb.2003.08.023}{\textit{Phys. Lett. B}~{\bf 572}, 133 (2003)}.

\bibitem{Jadach97} S. Jadach, W. P\l{}aczek, B.F.L. Ward, 
    \href{https://link.aps.org/doi/10.1103/PhysRevD.56.6939}{\textit{Phys. Rev. D}~{\bf 56}, 6939 (1997)}. 

\bibitem{Gounaris00} G. J. Gounaris, J. Layssac, and  F. M. Renard, 
\href{https://link.aps.org/doi/10.1103/PhysRevD.61.073013}{\textit{Phys. Rev. D}~{\bf 61}, 073013 (2000)}.


\bibitem{Rahaman17} R. Rahaman and R. K. Singh, 
\href{https://doi.org/10.1140/epjc/s10052-017-5093-1}{\textit{Eur. Phys. J. C}~{\bf 77}, 521 (2017)}.

\bibitem{inan21} S. C. \.{I}nan and  A. V. Kisselev, 
\href{https://doi.org/10.1007/JHEP10(2021)121}{\textit{J. High Energ. Phys.}~{\bf 2021}, 121 (2021)}.


\bibitem{Spor22} S. Spor, E. Gurkanli, M. Köksal, 
    \href{https://doi.org/10.1016/j.nuclphysb.2022.115785}{\textit{Nucl. Phys. B}~{\bf 979}, 115785 (2022)}.


\bibitem{Alcaraz00} J. Alcaraz, M. A. Falag\'an, and E. S\'anchez, 
\href{https://link.aps.org/doi/10.1103/PhysRevD.61.075006}{\textit{Phys. Rev. D}~{\bf 61}, 075006 (2000)}.

\bibitem{Gounaris00_2} G. J. Gounaris, J. Layssac, and  F. M. Renard, \href{https://link.aps.org/doi/10.1103/PhysRevD.62.073013}{\textit{Phys. Rev. D}~{\bf 62}, 073013 (2000)}.

\bibitem{ATLAS13} G. Aad {\it et al.} (ATLAS Collaboration), 
    \href{https://doi.org/10.1007/JHEP03(2013)128}{\textit{J. High Energ. Phys.}~{\bf 2013},128 (2013)}.


\bibitem{Demirci16} M. Demirci and A. I. Ahmadov, 
    \href{http://dx.doi.org/10.1103/PhysRevD.94.075025}{\textit{Phys. Rev. D}~{\bf 94}, 075025 (2016)}.


\bibitem{Demirci19b} M.~Demirci, 
    \href{http://link.aps.org/doi/10.1103/PhysRevD.100.075006}{\textit{Phys. Rev. D}~{\bf 100}, 075006 (2019)}.

\bibitem{Demirci20} M.~Demirci, 
    \href{https://doi.org/10.1016/j.nuclphysb.2020.115235}{\textit{Nucl. Phys. B}~{\bf 961}, 115235 (2020)}.


\bibitem{Demirci21} M.~Demirci and M. F. Mustamin, 
    \href{https://link.aps.org/doi/10.1103/PhysRevD.103.113004}{\textit{Phys. Rev. D}~{\bf 103}, 113004 (2021)}.

\bibitem{Juan08} J.J. Su, W.G. Ma, R. Y. Zhang, S.M. Wang, and L. Guo, 
    \href{https://link.aps.org/doi/10.1103/PhysRevD.78.016007}{\textit{Phys. Rev. D}~{\bf 78}, 016007 (2008)}.


\bibitem{Boudjema10} F. Boudjema, Le D. Ninh, S. Hao, and M. M. Weber, 
    \href{https://link.aps.org/doi/10.1103/PhysRevD.81.073007}{\textit{Phys. Rev. D}~{\bf 81}, 073007 (2010)}.


\bibitem{Heinemeyer16a} S. Heinemeyer and  C. Schappacher, 
    \href{https://doi.org/10.1140/epjc/s10052-016-4038-4}{\textit{Eur. Phys. J. C}~{\bf 76}, 220 (2016)}.

\bibitem{Heinemeyer18} S. Heinemeyer and  C. Schappacher, 
    \href{https://doi.org/10.1140/epjc/s10052-018-6009-4}{\textit{Eur. Phys. J. C}~{\bf 78}, 536 (2018)}.

\bibitem{HeYi22} L. He-Yi, Z. Ren-You , Ma Wen-Gan, J. Yi and L. Xiao-Zhou, 
    \href{https://doi.org/10.1088/1674-1137/ac424f}{\textit{Chinese Phys. C}~{\bf 46}, 043105 (2022)}.

\bibitem{Feynarts1} J.~K\"ublbeck, M. B\"ohm, and A. Denner, 
    \href{http://dx.doi.org/10.1016/0010-4655(90)90001-H}{\textit{Comput. Phys. Commun.}~{\bf 60}, 165 (1990)}.

\bibitem{Feynarts2} T.~Hahn, 
     \href{http://dx.doi.org/10.1016/S0010-4655(01)00290-9}{\textit{Comput. Phys. Commun.}~{\bf 140}, 418 (2001)}.

\bibitem{loop} T.~Hahn and M. Perez-Victoria, 
    \href{http://dx.doi.org/10.1016/S0010-4655(98)00173-8}{Comput. Phys.} \href{http://dx.doi.org/10.1016/S0010-4655(98)00173-8}{Commun.~{\bf 118}, 153 (1999)}.

\bibitem{CUBA} T. Hahn, 
    \href{https://doi.org/10.1016/j.cpc.2005.01.010}{\textit{Comput. Phys. Commun.}~{\bf 168}, 78-95 (2005)}.

\bibitem{Whizard} W. Kilian, T. Ohl, and J. Reuter, 
    \href{https://doi.org/10.1140/epjc/s10052-011-1742-y}{\textit{Eur. Phys. J. C}~{\bf 71}, 1742 (2011)}.
\bibitem{Omega} M. Moretti, T. Ohl, and J. Reuter, O'Mega: An Optimizing matrix element generator, LC-TOOL-2001-040-rev,
    \href{https://arxiv.org/abs/hep-ph/0102195}{arXiv:hep-ph/0102195-rev}.

\bibitem{CalcHep} A. Belyaev, N. D. Christensen, and A. Pukhov,
    \href{https://doi.org/10.1016/j.cpc.2013.01.014}{\textit{Comput. Phys. Commun.}~{\bf 184}, 1729 (2013)}.

\bibitem{Denner93} A. Denner, 
    \href{https://doi.org/10.1002/prop.2190410402}{\textit{Fortschr. Phys.}~{\bf 41}, 307 (1993)}.

\bibitem{Hooft72} G. 't Hooft and M. Veltman,
    \href{https://doi.org/10.1016/0550-3213(72)90279-9}{\textit{Nucl. Phys. B}~{\bf 44}, 189 (1972)}.


\bibitem{Kinoshita62}  T. Kinoshita, 
    \href{https://doi.org/10.1063/1.1724268}{\textit{J. Math. Phys.}~{\bf 3}, 650 (1962)}.

\bibitem{Lee64} T. D. Lee and M. Nauenberg, 
    \href{https://doi.org/10.1103/PhysRev.133.B1549}{\textit{Phys. Rev.}~{\bf 133}, 1549 (1964)}.

\bibitem{Schwinger49} J. Schwinger, 
    \href{https://link.aps.org/doi/10.1103/PhysRev.76.790}{\textit{Phys. Rev.}~{\bf 76}, 790-817 (1949)}.


\bibitem{PSS1} K. Fabricius, G. Kramer, G. Schierholz and I. Schmitt, 
    \href{https://doi.org/10.1007/BF01578281}{\textit{Zeit. Phys. C}~{\bf 11}, 315 (1982)}.
\bibitem{PSS2} G. Kramer and B. Lampe, 
    \href{https://doi.org/10.1002/prop.2190370302}{\textit{Fortschr. Phys.}~{\bf  37}, 161 (1989)}.
\bibitem{PSS3} H. Baer, J. Ohnemus and J. F. Owens, 
    \href{https://doi.org/10.1103/PhysRevD.40.2844}{\textit{Phys. Rev. D}~{\bf 40}, 2844 (1989)}.

\bibitem{PSS4} B. W. Harris and J. F. Owens, 
    \href{https://doi.org/10.1103/PhysRevD.65.094032}{\textit{Phys. Rev. D}~{\bf 65}, 094032 (2002)}.

\bibitem{Hooft79} G. 't Hooft and M. Veltman, 
    \href{https://doi.org/10.1016/0550-3213(79)90605-9}{\textit{Nucl. Phys. B}~{\bf 153}, 365 (1979)}.

\bibitem{PSS5} U. Baur, S. Keller, and D. Wackeroth, 
    \href{https://link.aps.org/doi/10.1103/PhysRevD.59.013002}{\textit{Phys. Rev. D}~{\bf 59}, 013002 (1998)}.

\bibitem{Omori99} T. Omori, {\it A Polarized positron beam for linear colliders}, 1st ACFA Workshop on Physics Detector at the Linear Collider,
    \href{https://inspirehep.net/literature/497328}{KEK-PREPRINT-98-237}.


\bibitem{Hikasa86} K. I. Hikasa, 
    \href{https://doi.org/10.1103/PhysRevD.33.3203}{\textit{Phys. Rev. D}~{\bf 33}, 3203 (1986)}.


\bibitem{PDG20} P. A. Zyla {\it et al.} (Particle Data Group),
    \href{https://doi.org/10.1093/ptep/ptaa104}{\textit{Prog. Theor. Exp. Phys.} \textbf{ 2020},  083C01 (2020)}.

\bibitem{Burkhardt95} H. Burkhardt and B. Pietrzyk, 
    \href{https://doi.org/10.1016/0370-2693(95)00820-B}{\textit{Phys. Lett. B}~{\bf 356}, 398 (1995)}.

\bibitem{Eidelman95} S. Eidelman and F. Jegerlehner, 
    \href{https://doi.org/10.1007/BF01553984}{\textit{Z. Phys. C}~{\bf 67}, 585 (1995)}.


\bibitem{Sirlin80} A.~Sirlin, 
    \href{https://link.aps.org/doi/10.1103/PhysRevD.22.971}{\textit{Phys. Rev. D}~{\bf 22}, 971 (1980)}.


\bibitem{Denner20} A. Denner and S. Dittmaier, 
    \href{https://doi.org/10.1016/j.physrep.2020.04.001}{\textit{Phys. Rep.}~{\bf 864}, 1-163 (2020)}.

\bibitem{Gounaris00_3} G. J. Gounaris, J. Layssac, P. I. Porfyriadis, and F. M. Renard, 
    \href{https://doi.org/10.1007/s100520000307}{\textit{Eur. Phys. J. C}~{\bf 13}, 79–97 (2000)}.

\bibitem{Bardin17} D. Bardin, S. Bondarenko, P. Christova, L. Kalinovskaya, W. von Schlippe, and E. Uglov, 
    \href{https://doi.org/10.1134/S1547477117060061}{\textit{Phys. Part. Nuclei Lett.}~{\bf 14}, 811–816 (2017)}.

\end{thebibliography}
\end{document}